\documentclass[useAMS,usenatbib,usegraphicx]{mn2e}
\usepackage{epsfig}
\usepackage{amsmath} 
\usepackage{rotating}           
\usepackage{color}     
\usepackage{graphicx}
\usepackage{times}
\usepackage{upgreek} 

\def\kms {\hbox{${\rm km\ s}^{-1}$}}
\def \HI {H{\sc \,i}}

\def\MOLH {\hbox{${\rm H}_2$}}  

\def\lapp{\ifmmode\stackrel{<}{_{\sim}}\else$\stackrel{<}{_{\sim}}$\fi}
\def\gapp{\ifmmode\stackrel{>}{_{\sim}}\else$\stackrel{>}{_{\sim}}$\fi}

\title[The scarcity of OH absorption]{On the scarcity of redshifted OH and millimetre-band molecular absorption}

\author[S. J. Curran]{S. J. Curran\thanks{Stephen.Curran@vuw.ac.nz}\\
School of Chemical and Physical Sciences, Victoria University of Wellington, PO Box 600, Wellington 6140, New Zealand}

\begin{document}

 \date{Accepted ---. Received ---; in original form ---}

\pagerange{\pageref{firstpage}--\pageref{lastpage}} \pubyear{2021}

\maketitle

\label{firstpage}

\begin{abstract}
  Despite much searching, redshifted decimetre and millimetre-band absorption by molecular gas remains very rare,
  limited to just six systems at $z_{\rm abs} \gapp0.05$. Detection of these transitions can yield precise diagnostics
  of the conditions of the star forming gas in the earlier Universe, the hydroxyl (OH) radical being of particular interest
  as in the $\lambda = 18$~cm ground state there are four different transitions located close to \HI\ 21-cm and thus
  detectable with the Square Kilometre Array and its pathfinders. The four transitions of  OH have very different
  dependences on the fundamental constants, thus having much potential in testing for any evolution in these over large
  look-back times.  By collating the photometry in a uniform manner, we confirm our previous hypothesis that the
  normalised OH absorption strength is correlated with the optical--near-infrared red colour of the sight-line. Applying
  this to the published searches, we find that all, but one (J0414+054), have simply not been searched sufficiently deeply. 
   We suggest that
  this is due to the standard  selection of sources with reliable optical redshifts introducing a bias against those with enough
  dust with which to shield the molecular gas. For the single source searched to sufficient depth, we have reason to
  suspect that the high degree of reddening arises from another system along the sight-line, thus not being inconsistent
  with our hypothesis. We also show that the same optical redshift bias can account for the scarcity of millimetre-band
  absorption. 
\end{abstract}
\begin{keywords} 
galaxies: active -- quasars: absorption lines -- radio lines: galaxies  -- galaxies: high redshift -- techniques: photometric -- methods: data analysis
\end{keywords}

\section{Introduction} 
\label{intro}

Although neutral hydrogen (\HI) has been detected through $\lambda=21$~cm absorption in 142 redshifted ($z_{\rm abs}
\gapp0.1$) sources (\citealt{cur21a} and references therein), absorption by the hydroxyl (OH) radical at $\lambda=18$~cm
remains rare, with only five detections at these redshifts \citep{cdn99,kc02a,kcdn03,kcl+05}.  
As well as tracing the molecular, and hence cool
star-forming, gas, OH in this band is of particular interest as there are four hyperfine transitions (at 1612, 1665, 1667
\& 1720~MHz), each of which has a different dependence on various combinations of the fundamental constants (fine
structure, electron--proton mass ratio, proton g-factor). The main advantage is that the absorption is known to arise 
along the same sight-line \citep{dar03}, thus bypassing the possible line-of-sight differences which plague optical-band
data \citep{mwfc03} and  could therefore contribute to the observed evolution in the fine structure constant  over the past 12
billion years (\citealt{wkm+10} and references therein). Furthermore, the frequency shifts of the of radio-band lines are an order 
of magnitude more sensitive to a change in the fine structure constant than the optical-band lines \citep{cdk04}.

Despite much searching, however, redshifted OH absorption remains scarce. As a possible explanation, \citet{cwm+06}
noted that each of the detections arose along a very red sight-line, with an optical--near-infrared colour of
$V-K\gapp5$, whereas the non-detections tended to be bluer. Although \MOLH\ has been detected in the rest-frame ultraviolet-band
in $\approx30$ damped Lyman-$\alpha$ absorption systems (DLAs, e.g. \citealt{lv85,bki+14,nsr+15}), with colours
$V-K\lapp4$, these have  much lower molecular fractions than detectable in OH \citep{cwc+11}.  This suggests that the
selection of optically bright sources introduces a bias against dustier, and more molecular rich, sight-lines.  Since
then there have been many more searches for OH absorption
\citep{cdbw07,cww+08,cwm+10,caw+16,amm+16,cwa+17,gdb+15,gdlb20,dsgj19,zls+20,gss+21}, resulting in only one new detection (at
$z_{\rm abs} = 0.05$, \citealt{gms+18}).

Although we have previously noted a possible correlation between the OH absorption strength and the
optical--near-infrared colour of the sight-line, this is rarely considered in the discussion of the low detection rates
(e.g. \citealt{gms+18,zls+20}).  Given that surveys for decimetre-band absorption lines (\HI\ \& OH) are currently
underway on the pathfinder telescopes of the {\em Square Kilometre Array} (SKA)\footnote{Specifically, the {\em First
    Large Absorption Survey in \HI} on the {\em Australian Square Kilometre Array Pathfinder} (\citealt{asd+19}) and the
  {\em MeerKAT Absorption Line Survey} \citep{gjs+21}}, we revisit our hypothesis by adding the results of the 
OH searches made since. In order to be applicable to future large surveys, we  formalise the 
collation of the magnitude measures, obtaining these in an automated and more uniform manner than previously.  
We then normalise the OH limits according to our suspicion that the line width is close to that of the \HI\ 21-cm,
before discussing the limits in the context of the detected OH and millimetre-band absorption systems.

\section{Analysis}

\subsection{OH searches}

In the sixteen years since the last published OH detection, there has only been one further detection, at a redshift of
$z_{\rm abs} = 0.05$ \citep{gms+18}.  Four of these six redshifted absorbers arise in galaxies intervening the
sight-line to a more distant source providing the radio continuum, with the remaining two associated with the continuum
source itself.  Furthermore, for the OH absorbers we note that:
\begin{itemize}
  \item[--] All were previously detected in \HI\ 21-cm absorption \citep{cps92,cry93,cmr+98,cdn99,kb03,gms+18}.
       \item[--] At the time, at least five were located along very red sight-lines, with optical--near-infrared colours of $V-K\gapp5$, which suggested that the reddening was due to dust which protected the molecules from the ambient ultra-violet field \citep{cwm+06}.
        \item[--] The same five  also exhibit molecular absorption in the millimetre-bands \citep{wc94,wc95,wc96,wc96b,wck18},
          four of which were detected in these transitions prior to the OH absorption and one prior to the \HI\ absorption.
\end{itemize}
We therefore consider the OH searches to date in which \HI\ 21-cm absorption has been detected\footnote{In the absence of an a priori redshift, spectral scans in the millimetre-band are very observationally expensive, due to the narrow relative bandwidths at $\nu\gapp100$~GHz \citep{mcw03}.},
which we list in Table~\ref{sources}.
\begin{table*}  
\centering
\begin{minipage}{170mm}
  \caption{The published OH 18-cm searches for which \HI\ 21-cm absorption has been detected. The type (intervening or
associated) is followed by the integrated optical depth [\kms] and the full-width at half maximum of the profile [\kms]
for both the \HI\ and 1667~MHz OH absorption. 
Where OH is not detected, the spectral resolution which gives the quoted $3\sigma$ upper limit is given. 
See Sect.~\ref{sec:colours} for the $V$ and $K$ magnitudes.}
\begin{tabular}{@{}l c c   ccc   rcc  r c@{}} 
\hline
\smallskip
Sight-line & $z_{\rm abs}$ & Type &  \multicolumn{3}{c}{Hydrogen} & \multicolumn{3}{c}{Hydroxyl}  & $V$ & $K$ \\
          &                                    &         &  $\int\tau_{\text{\HI}}dv$ & FWHM$_{\text{\HI}}$ & Ref. & $\int\tau_{\rm OH}dv$ & FWHM$_{\text{OH}}, \Delta v$ & Ref. & & \\
\hline
CGRaBS\,J0111+3906             &  0.66847  &A &  43.7  & 94  &O06 & $<0.84$ &  19   & C10 & --- & 15.07\\
PKS\,0132--097       		  & 0.76335  &I &    7.06  &   151 &  K03a & 4.00  &  130  & K05 & 22.55 &  13.58 \\
B2\,0218+35	                     & 0.688466 & I& 2.65  &   53  &   K03b   & 0.401  &  61 & K03b & 21.92 & 14.85\\
\protect[HB89]\,0248+430    	 & 0.0519  & I & 0.42    &  21    & G18a & 0.08    &  19  & G18a & --- &  14.56\\
4C\,+05.19  (J0414+054)	   & 0.95974 & I&   0.88  &  49  & C07&  $<0.02$  & 4.3 & C07 & 24.14 &  13.77\\
PKS\,0428+20                          & 0.219      & A &   1.90  & 475 &  V03 &  $<0.10$  &  85.5& O05 & $>20.0$  &  14.47 \\
PKS\,0500+019                        & 0.58467  &A&   3.4  &     100  & C98 & $<0.23$ &   18 & C06 & $>21.0$   & 15.56 \\
B3\,0839+458                         	 & 0.1919 &A	 & 25.9	& 79.2  &M17 &   $<0.07$ &1.64 & Z20 & 19.77 &14.33 \\
SDSS\,J084957.97+510829.0  &0.3120 & I & 0.95& 11&G13& $<0.091$  & 0.9& G18a & --- & 13.97\\
WISE\,J085042.24+515911.6    & 1.3265  & I & 15.3 & 30 & G09& $<0.244$ & 1.1 & D20 &19.17&  15.26 \\
WISEA\,J085244.73+343540.6 & 1.3095 & I & 6.91 & 48 & G09 & $<0.092$ & 0.4 & D20 &  19.40 &15.08 \\
WISEA\,J090325.55+162256.0 & 0.1823 &A& 8.98 &    163	& M17 &  $<0.98$ &  1.62 & Z20 & 18.88 & 13.92 \\
3C\,216 (0906+430)             & 0.670  &A& 0.67 &  285 &  V03 & $< 0.024$ & 30	&  G19&  18.48 &  14.56\\
2MASS\,J09422198+0623362 & 0.12368  & A & 49.9	&  50   &S15	& $<0.070$&  3.3 &G18a & 19.51&  15.13\\
B3\,1036+464	                	&  0.1861 &A&  14.75 &86  & M17 &  $<0.11$ & 1.63 & Z20& --- & 14.77 \\
SDSS\,J104257.58+074850.5 &  0.0331& I & 0.19 & 3.6 &B10 	& $<0.027$ & 1.5	 & G18a & 18.97 &  ---\\
PKS\,1107--187                   &  0.48909 & A&  0.99	&  50   & C10 &  $<0.077$ &   10  & C10 & $>19.6$  & 15.23\\
\protect[HB89]\,1127--145 &  0.31272   & I & 3.41  & 47   & G20 & $<0.098$ &   30 & G20 & 19.51 & 15.13\\
2MASX\,J11233202+2350475&   0.2070  & A & 5.01 &     220 &  M17 &  $<0.25$ &   1.66 & Z20 & 19.35 &  13.95\\
B3\,1241+410                      &  	 0.0169	& I & 2.24 & 15& G18b	 & $<0.035$ & 1.4&G18a & 20.01 & ---\\
\protect[HB89]\,1243--072 &     0.436734  & I & 	1.50  & 4.4 &	G20	&  $<1.1$ &   30	&   G20 & 20.13 & --- \\
2MASS\,J12415751+6332417  & 0.143	 & I	&  2.90&   40 &  G10 & $<0.099$  &  1.2	&G18a & 18.20 &  15.14\\
WISEA\,J130132.60+463402.9 	&	 0.2055	& A&4.24 & 172  &M17 &  $<0.25$ &  1.65 & Z20 & 19.38 &  13.67\\
\protect[HB89]\,1413+135     & 0.24671& A& 7.1  &     18   & C92 &   0.023  &  8 & K02 & 21.15 & 14.86\\
WISEA\,J142210.83+210554.2 &	 0.1915 & A&  8.36 &180  & M17 &  $<0.31$ & 1.64 & Z20& --- & 14.84\\
B2\,1504+37                             &  0.67343  & A &24.5  & 57  & C98 &   0.42  &   45 &   K02 & 22.68 &  15.52\\
WISEA\,J153452.94+290919.9 	 &	 0.2010	 & A& 6.77 & 234  & M17	& $<0.60$ & 	1.65 & Z20 & 19.80 &  14.56\\
2MASX\,J17081524+2111175  & 0.2241&A&28.4	&  211  &   M17 &    $<1.59 $  &  1.68 & Z20 & 20.47 &  13.79\\
PKS\,1830--21	                      & 0.885316  & I & 5.5  & 199   & C99  & 1.33  &  239   & C99 & $>22.3$ & 15.38 \\
PKS\,2052+005                         &  0.20153& I & 29.9 & 150  & D16 & $<0.03$ & 1.64 & Z20 & 21.13 &  15.43 \\
PKS\,2252--089         & 0.6064   & A&  11.0 & 94& C10 &  $<1.1$ &   10	& C10 & 22.30 & 16.91\\
3C\,459	(2313+03)   &   0.2199& A & 0.388&  164  & V03 &$<0.13$& 30& G19 & 17.44 & 13.92\\
\hline
\end{tabular}
{References: C92 -- \citet{cps92}, C98 --- \citet{cmr+98}, C99 -- \citet{cdn99}, K02 -- \citet{kc02a}, K03a -- \citet{kb03}, K03b -- \citet{kcdn03},  V03 -- \citet{vpt+03}, K05 -- \citet{kcl+05}, O05 -- \citet{ogs+05}, O06 -- \citet{omd06}, C06 -- \citet{cwm+06}, C07 -- \citet{cdbw07}, G09 -- \citet{gsp+09a}, B10 -- \citet{bty+10}, C10 -- \citet{cwm+10}, G10 -- \citet{gsb+10}, G13 -- \citet{gsn+13}, S15 -- \citet{sgmv15}, D16 -- \citet{dsg+16}, M17 -- \citet{mmo+17}, G18a --  \citet{gms+18}, G18b -- \citet{gsf+18}, G19 -- \citet{gdb+15}, D20 -- \citet{dsgj19}, 
G20 -- \citet{gdlb20}, Z20 -- \citet{zls+20}.}
\label{sources}  
\end{minipage}
\end{table*}

\subsection{Colours}
\label{sec:colours}

Previously \citep{cwm+06,cdbw07,cww+08,cwm+10}, the $V-K$ colours came from a variety of disparate sources in the
literature. As well as being non-uniform, this is not ideal when considering a large number of sources, which
will be the case in the analysis of SKA and SKA pathfinder data. We therefore use a similar method to that of
\citet{cm19}, where the photometry are scraped from {\em NASA/IPAC Extragalactic
  Database} (NED), the {\em Wide-Field Infrared Survey Explorer} (WISE, \citealt{wem+10}) and the {\em Two Micron All Sky
 Survey} (2MASS, \citealt{scs+06}). If a datum fell into either the $V$ or $K$-bands, this was used,
with multiple values being averaged. If not, this was extrapolated from a power-law fit to near-by bands --
the WISE $W1$ \& $W2$-bands for $K$ and the $I,R$ \& $B$-bands for $V$ (Fig.~\ref{SEDs}).
\begin{figure*}
\centering 
\includegraphics[angle=270,scale=0.27]{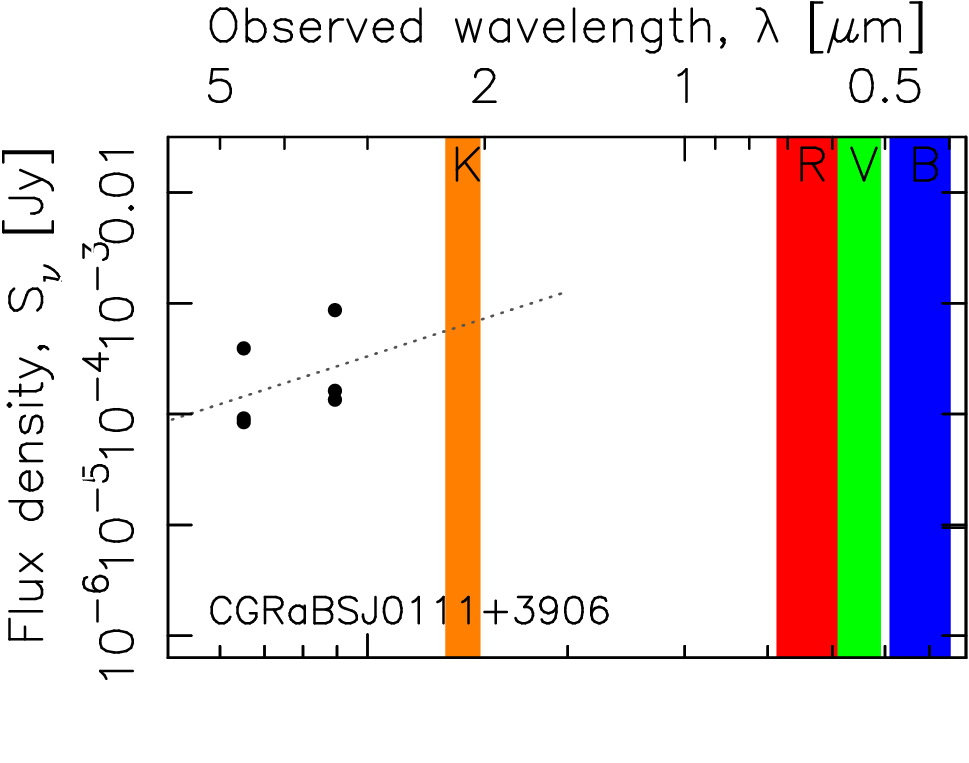}
\includegraphics[angle=270,scale=0.27]{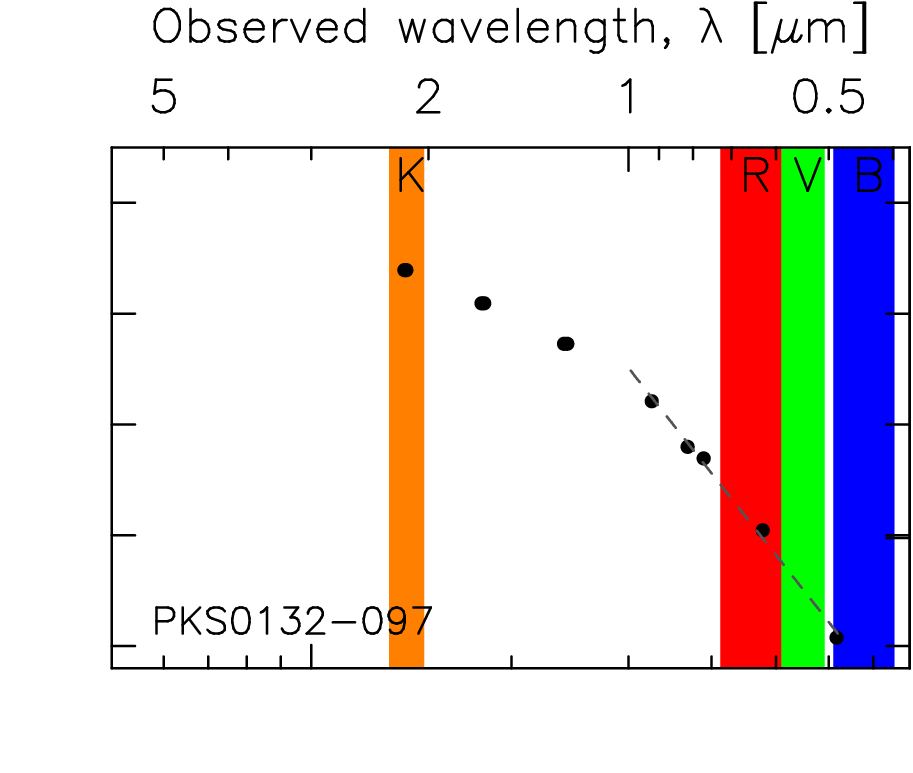}
\includegraphics[angle=270,scale=0.27]{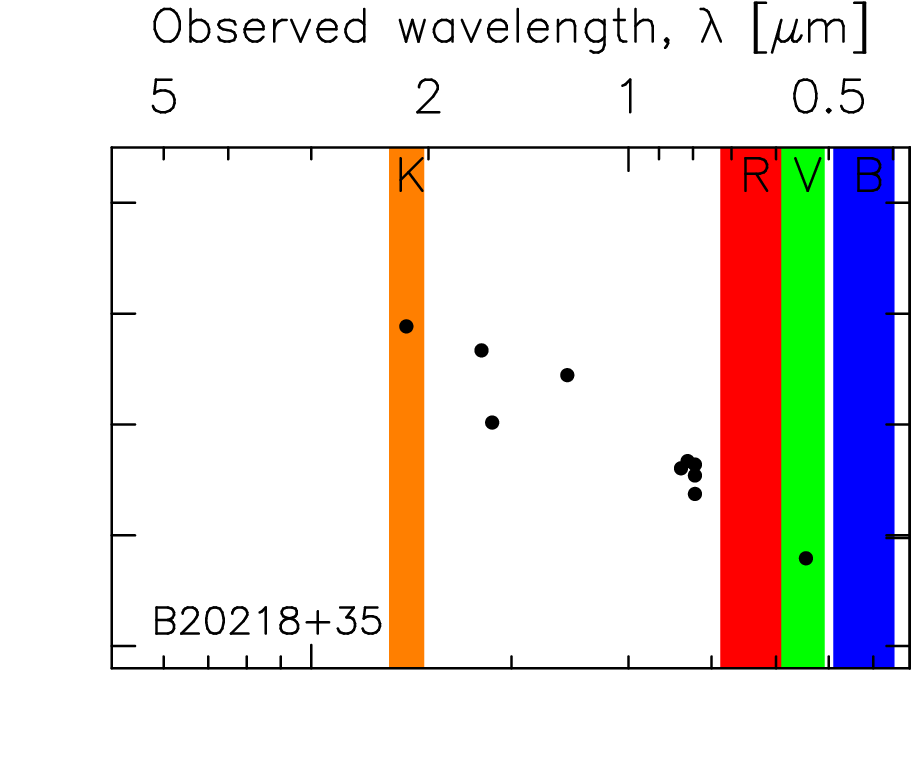}
\includegraphics[angle=270,scale=0.27]{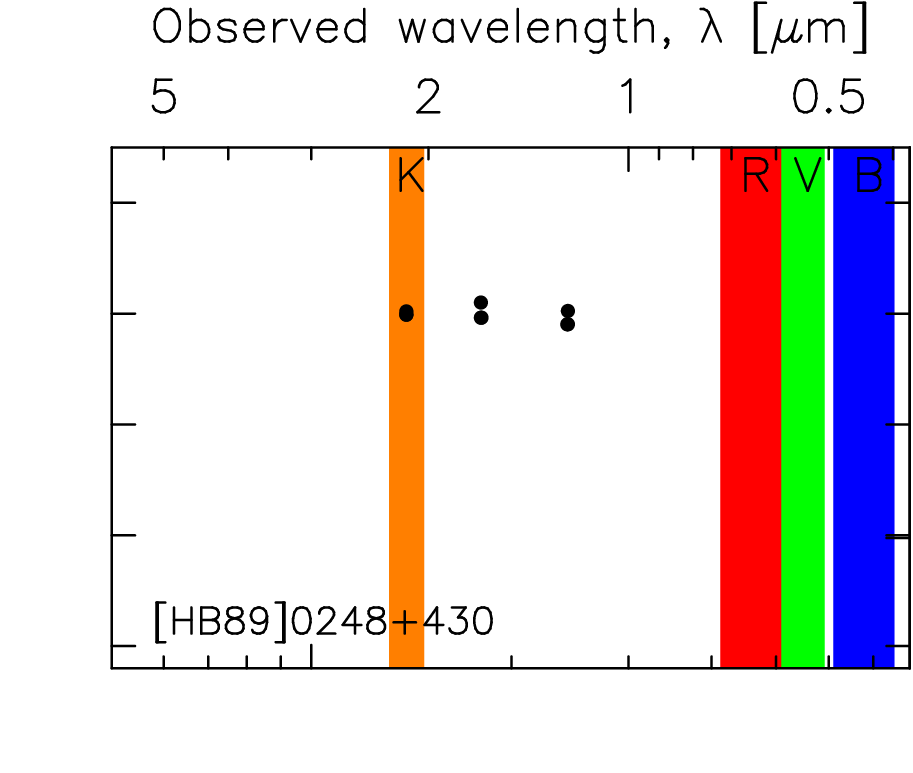}
\includegraphics[angle=270,scale=0.27]{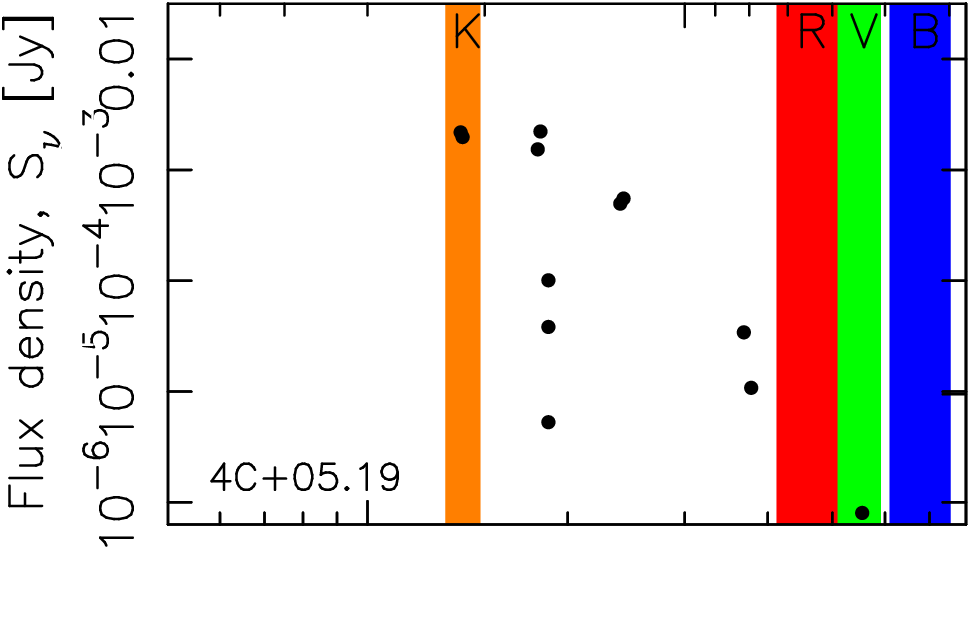}
\includegraphics[angle=270,scale=0.27]{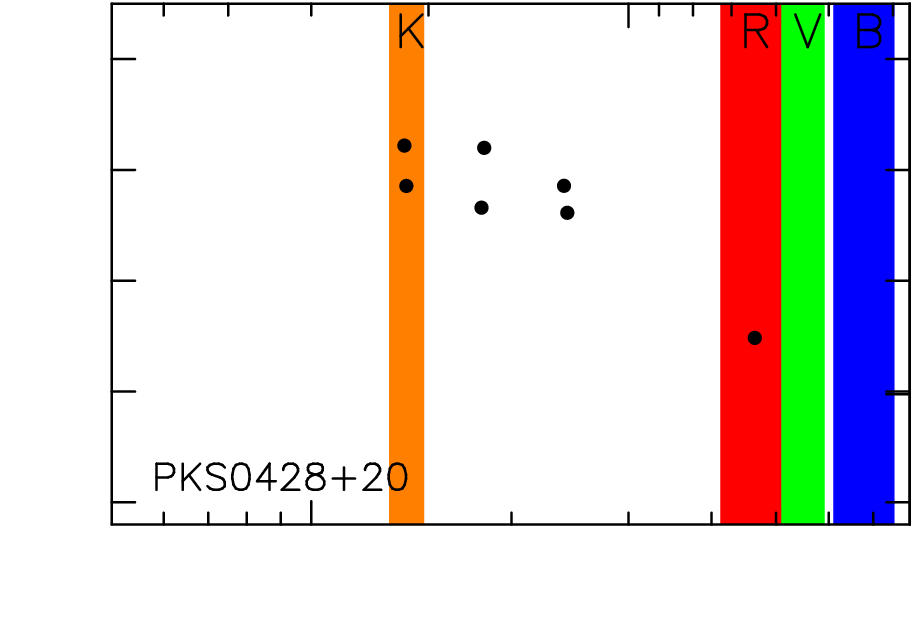}
\includegraphics[angle=270,scale=0.27]{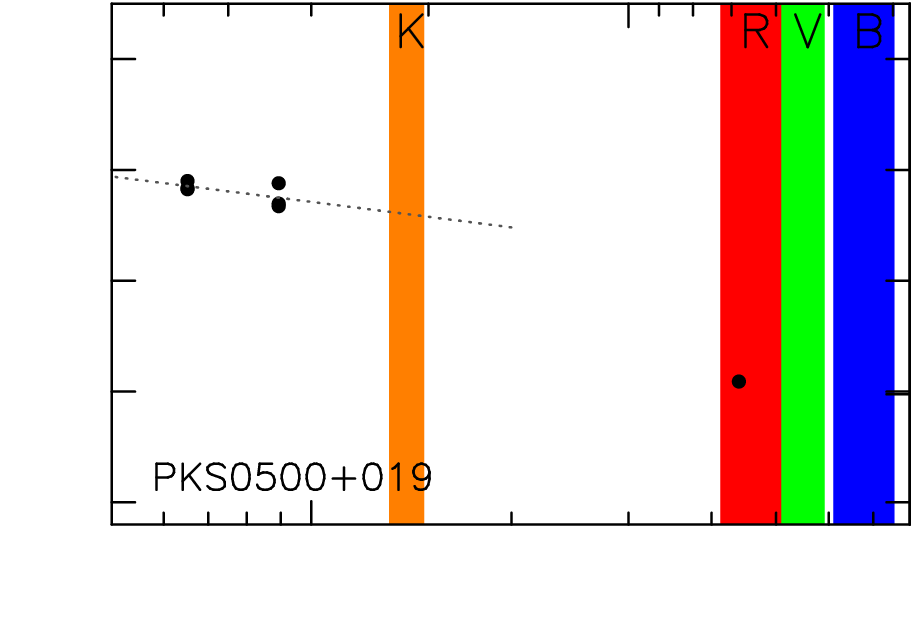}
\includegraphics[angle=270,scale=0.27]{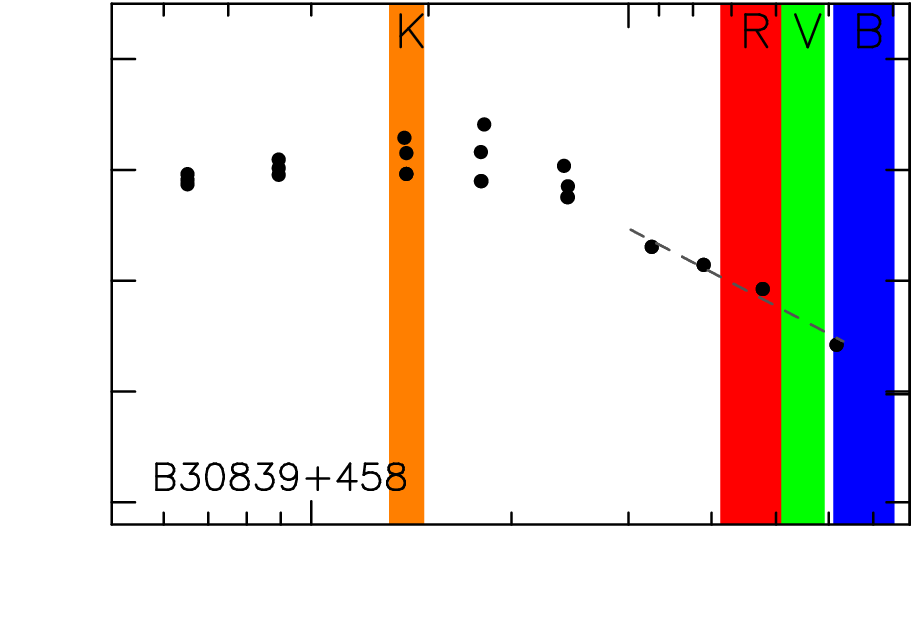}
\includegraphics[angle=270,scale=0.27]{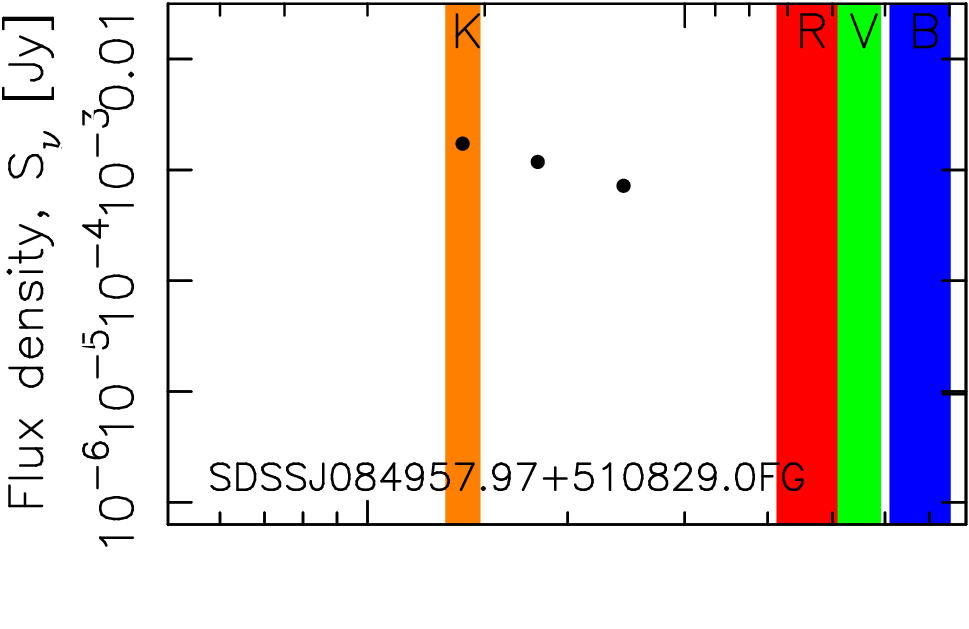}
\includegraphics[angle=270,scale=0.27]{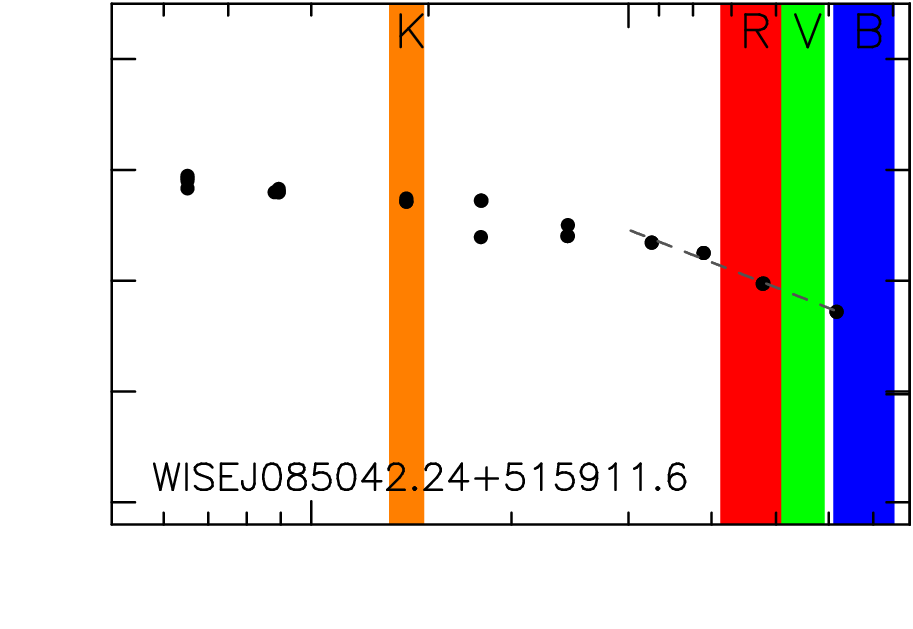}
\includegraphics[angle=270,scale=0.27]{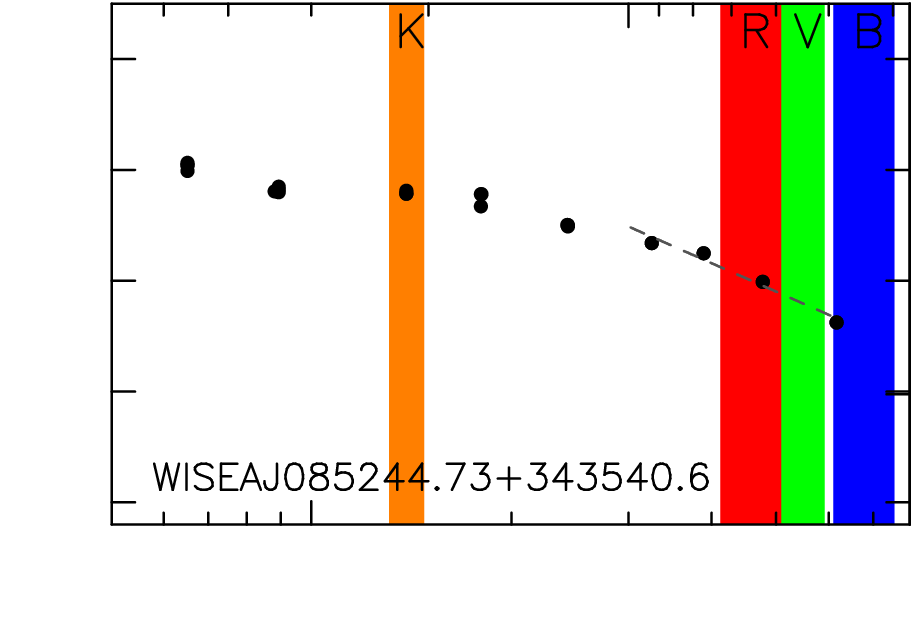}
\includegraphics[angle=270,scale=0.27]{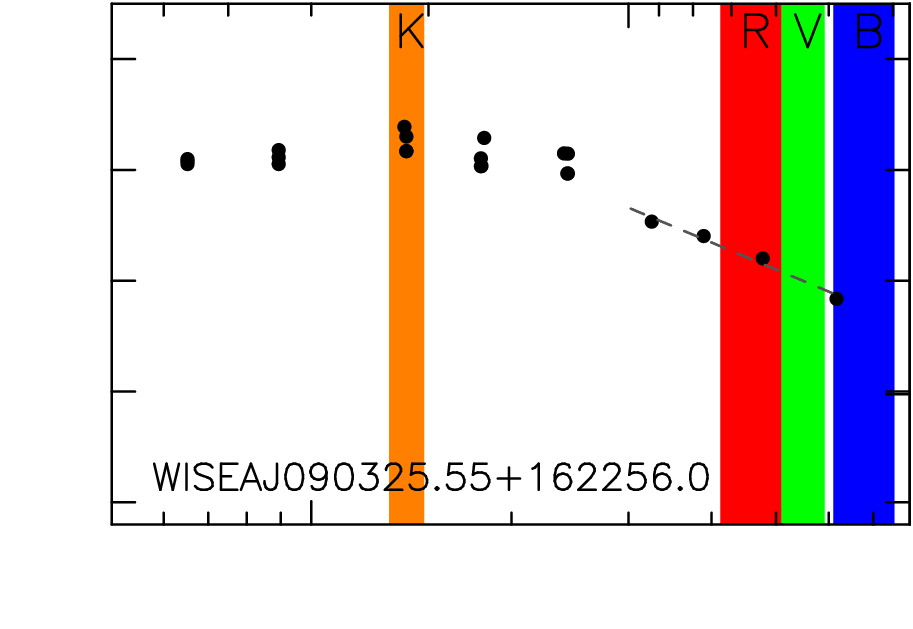}
\includegraphics[angle=270,scale=0.27]{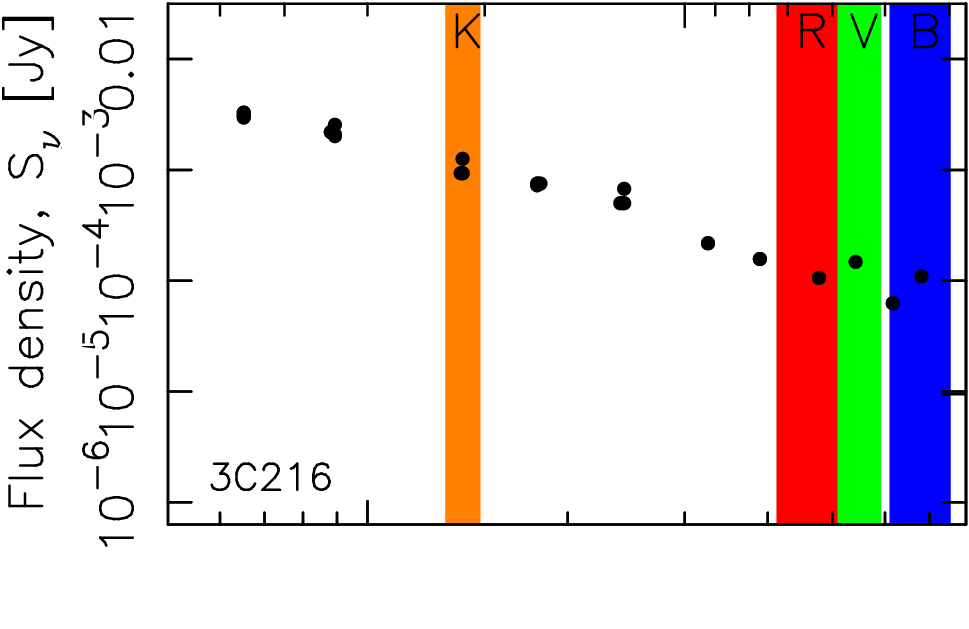}
\includegraphics[angle=270,scale=0.27]{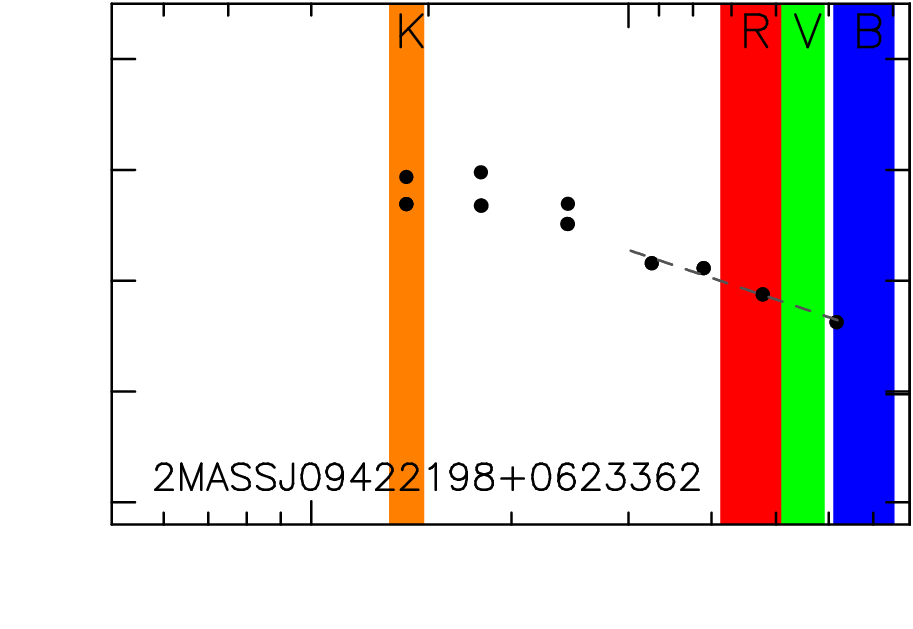}
\includegraphics[angle=270,scale=0.27]{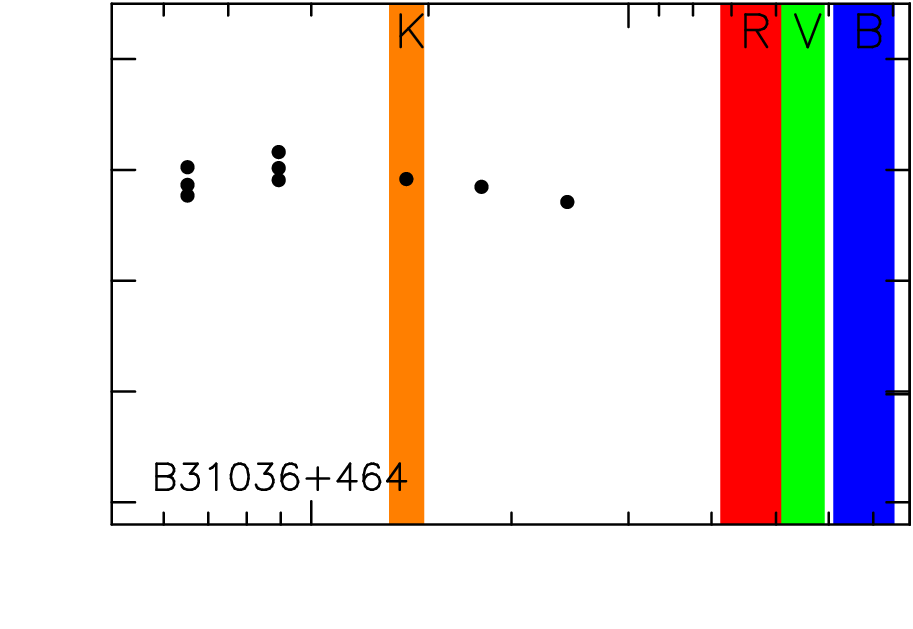}
\includegraphics[angle=270,scale=0.27]{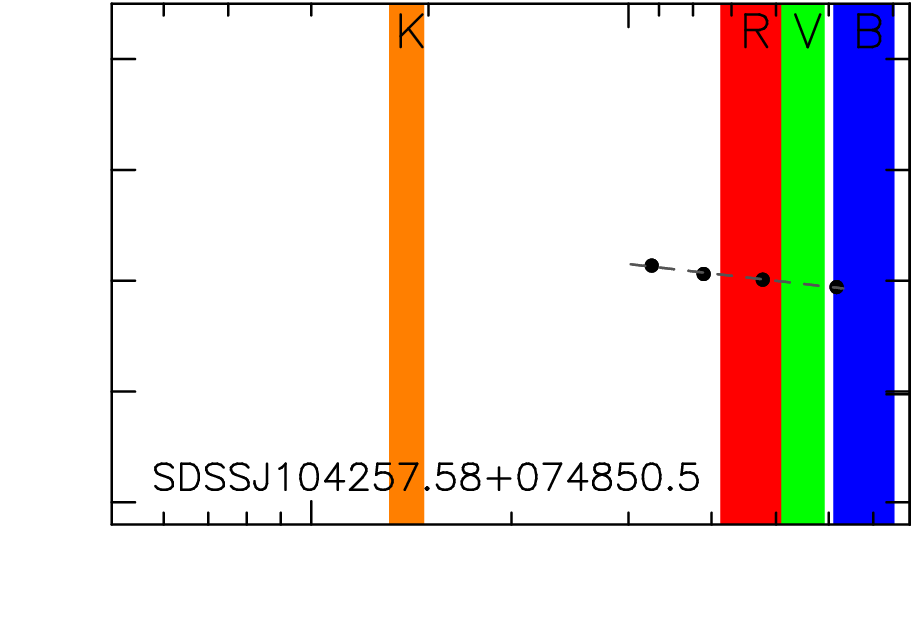}
\includegraphics[angle=270,scale=0.27]{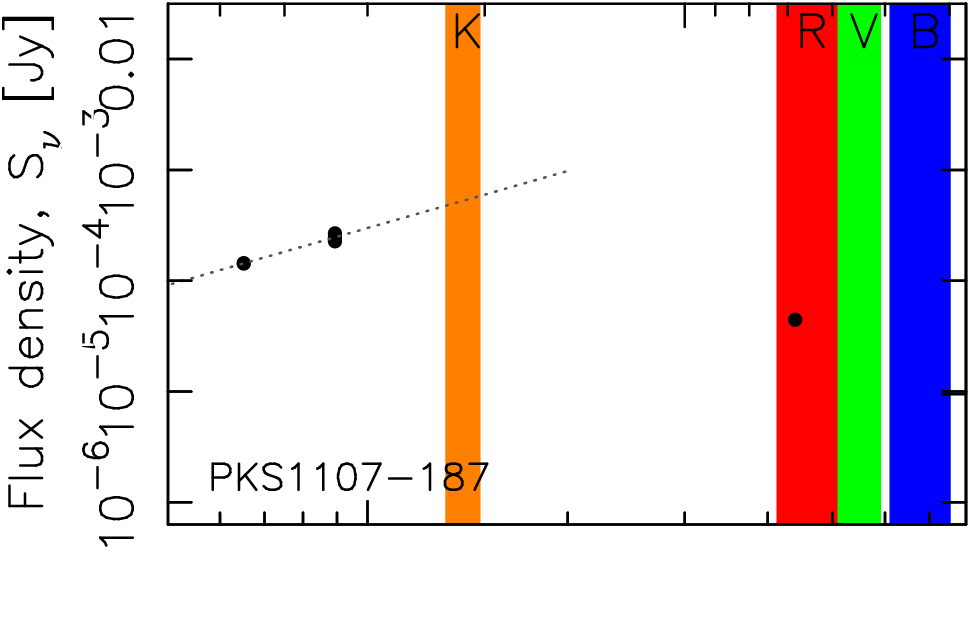}
\includegraphics[angle=270,scale=0.27]{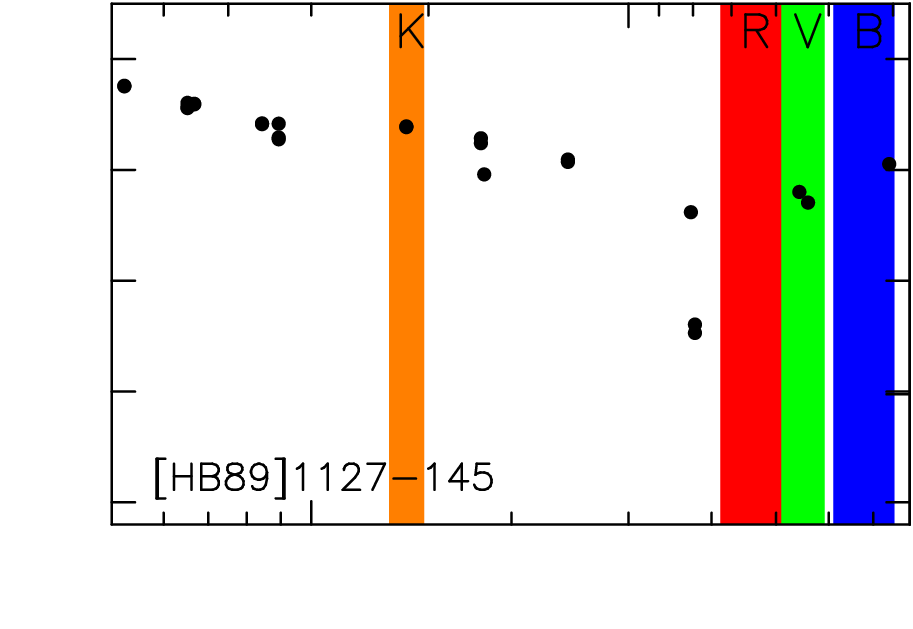}
\includegraphics[angle=270,scale=0.27]{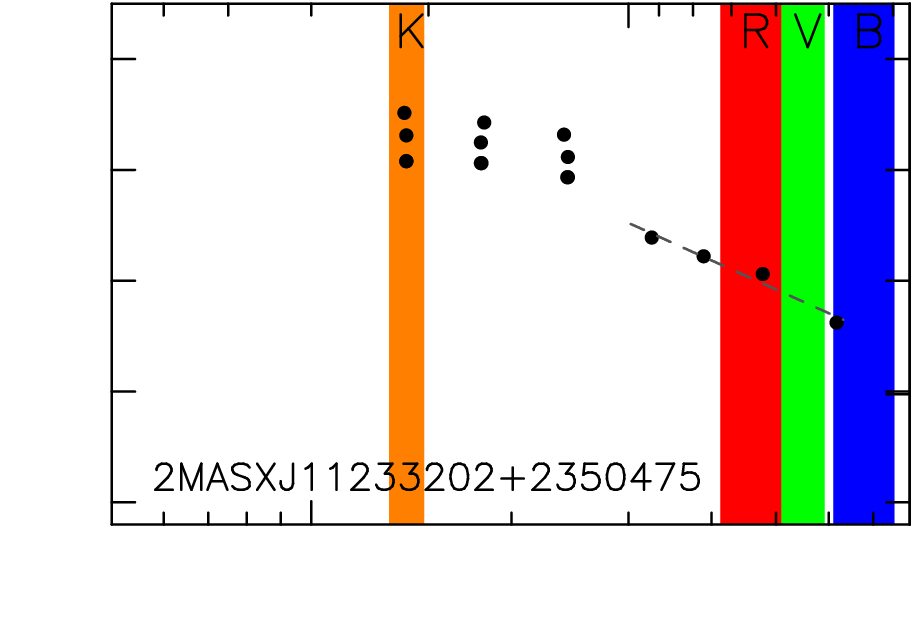}
\includegraphics[angle=270,scale=0.27]{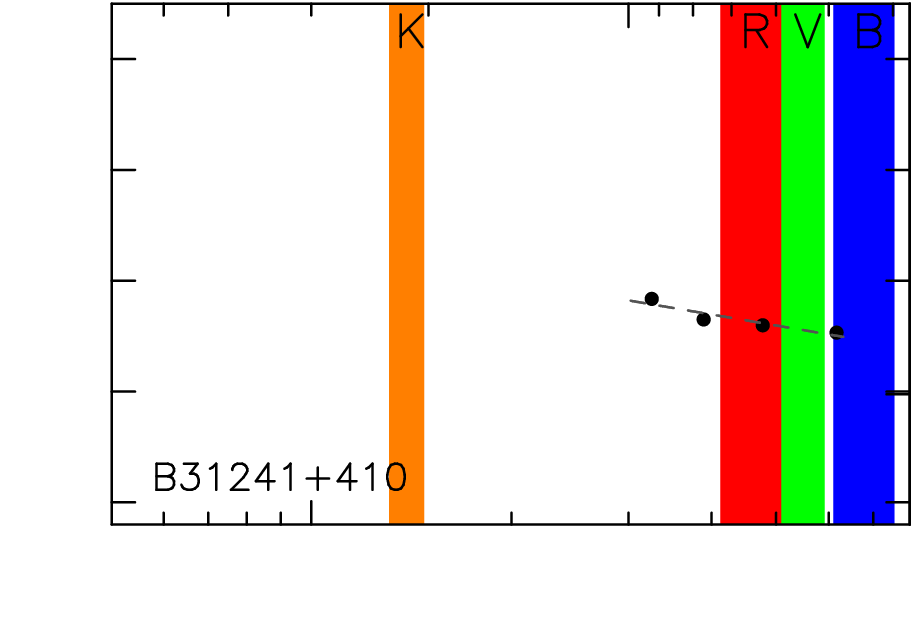}
\includegraphics[angle=270,scale=0.27]{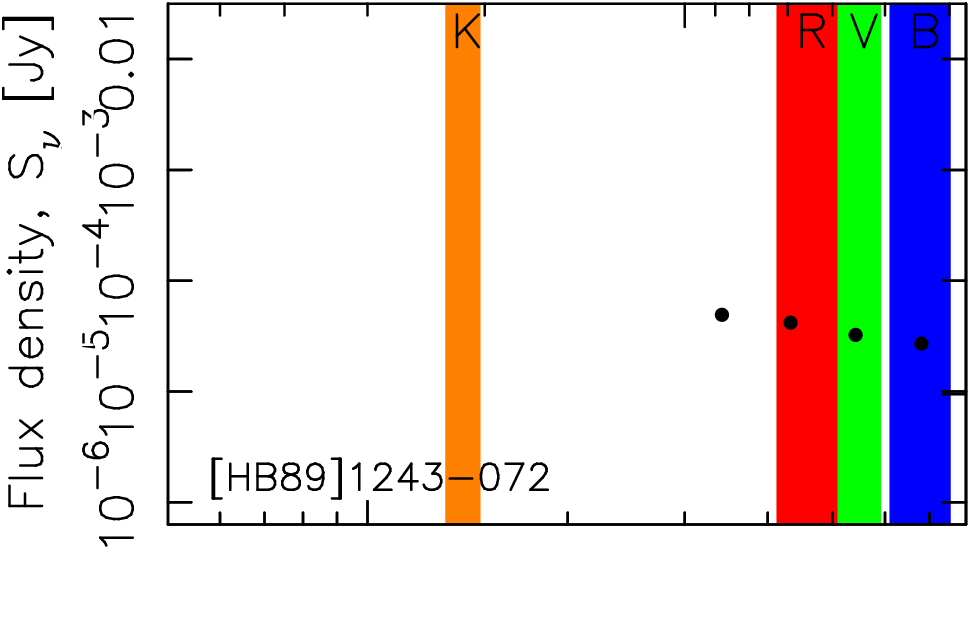}
\includegraphics[angle=270,scale=0.27]{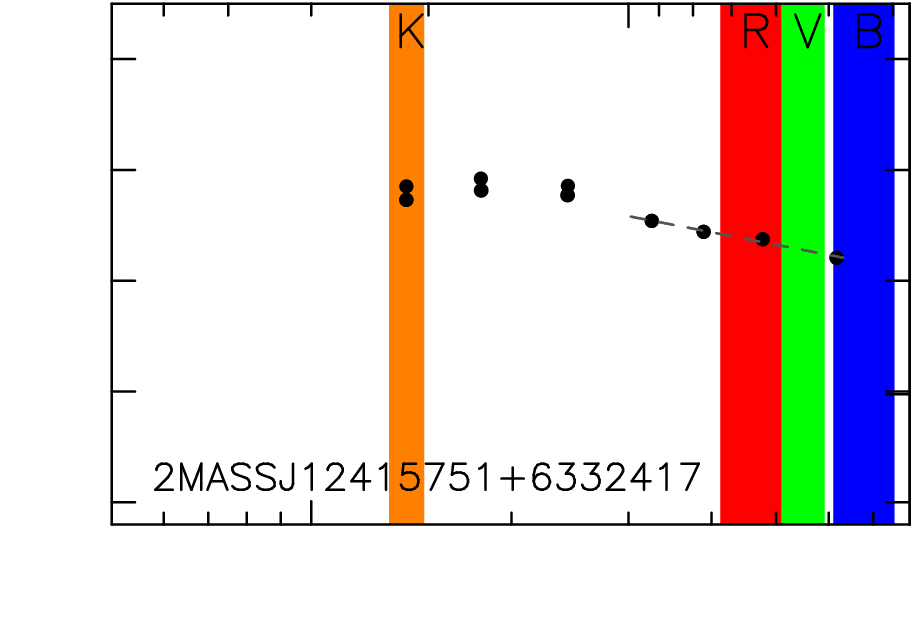}
\includegraphics[angle=270,scale=0.27]{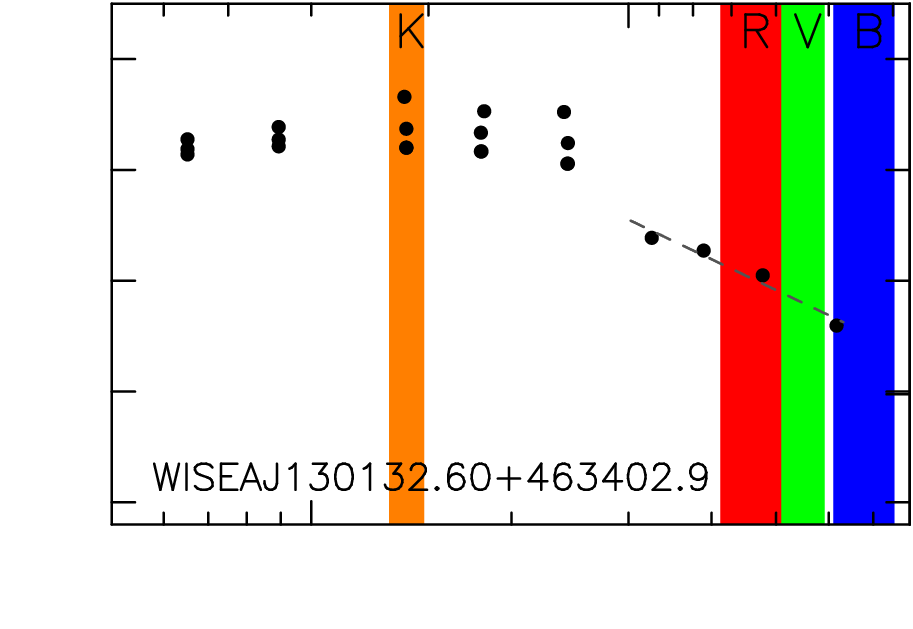}
\includegraphics[angle=270,scale=0.27]{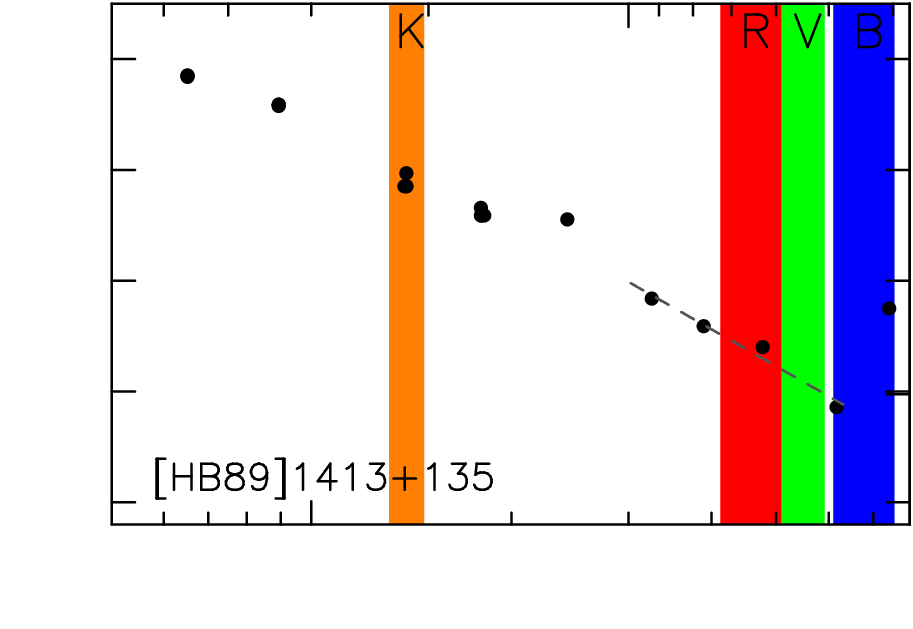}
\includegraphics[angle=270,scale=0.27]{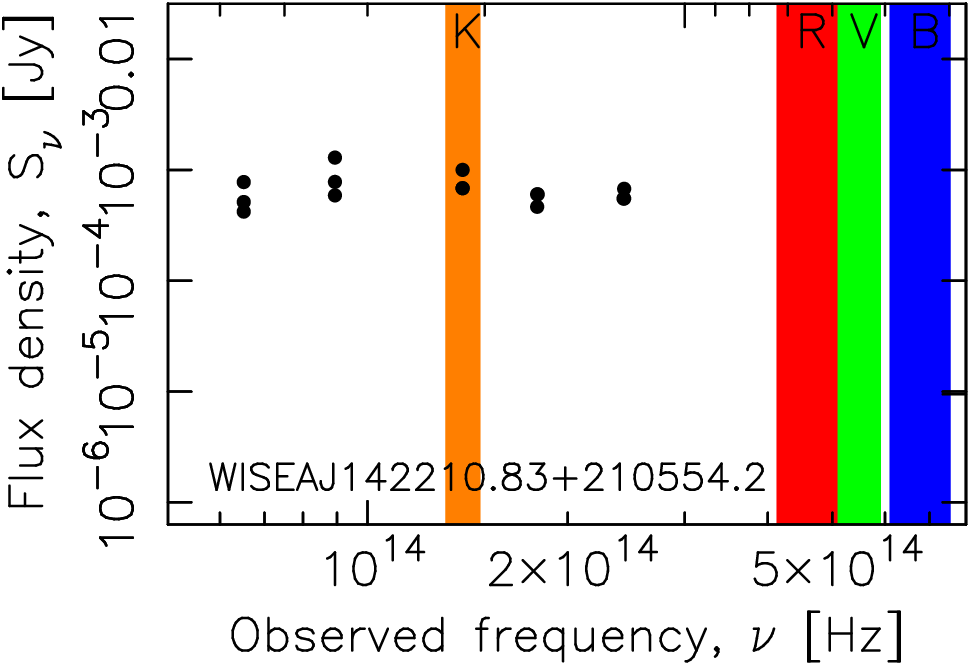}
\includegraphics[angle=270,scale=0.27]{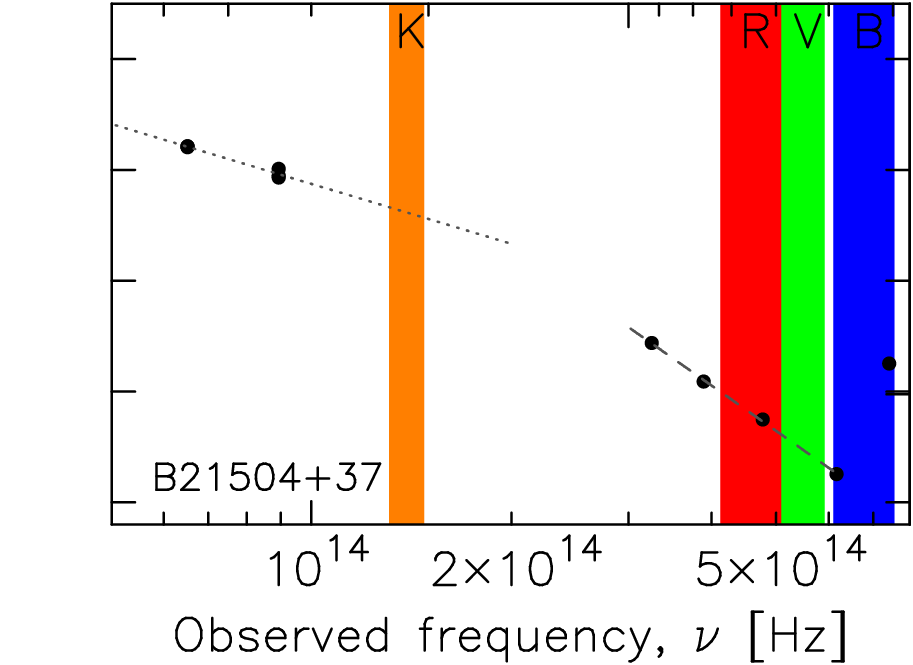}
\includegraphics[angle=270,scale=0.27]{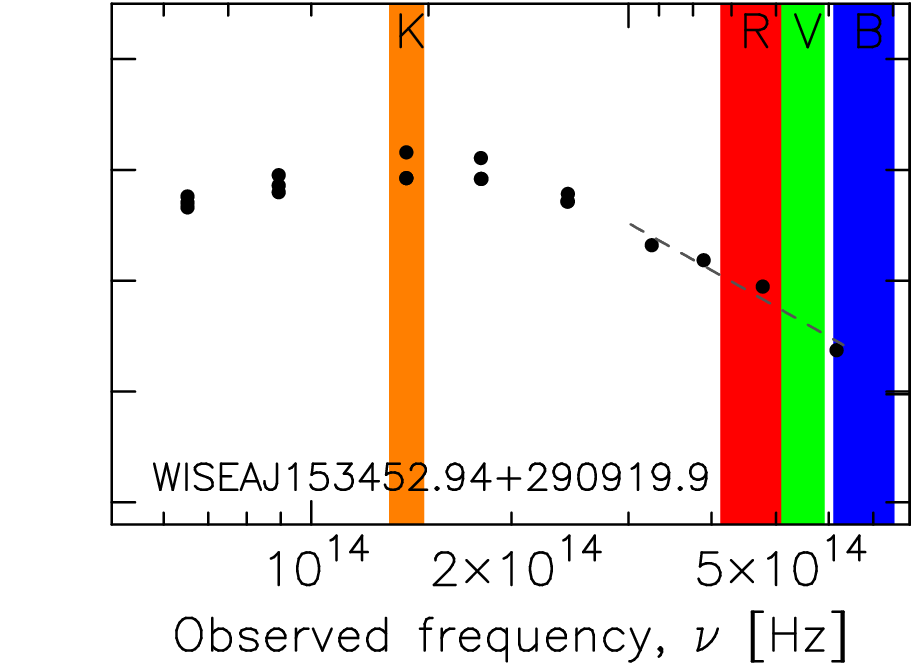}
\includegraphics[angle=270,scale=0.27]{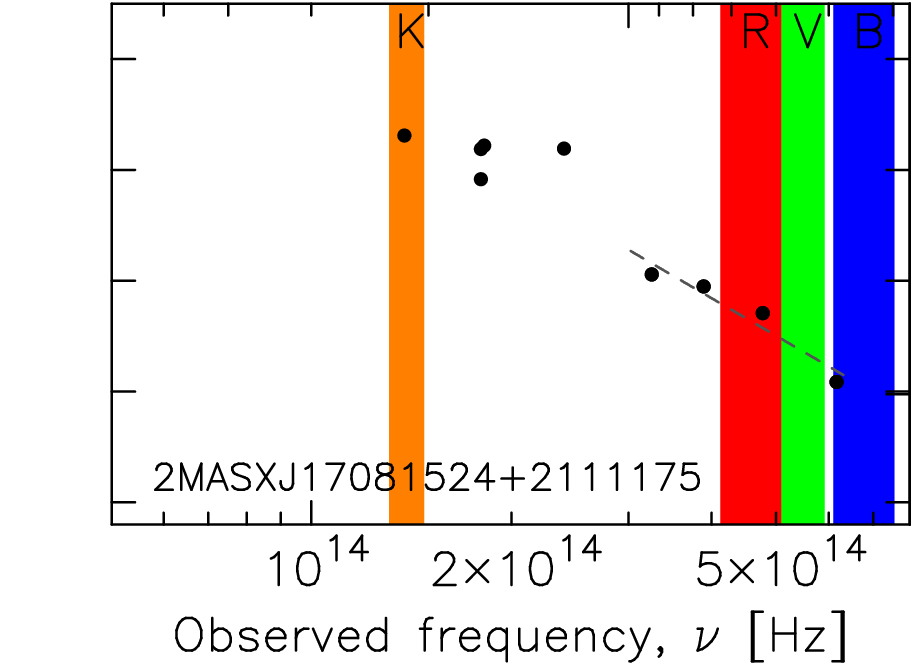}
\caption{The near-infrared--visible photometry for the sight-lines searched in OH and detected in \HI\ 21-cm absorption.
The coloured regions show the range of the $B,V,R$ and $K$-bands and the lines the least-square power-law fits
from the neighbouring photometry, in the cases where the $V$ and $K$-bands are empty 
(dotted -- near-infrared \& broken -- optical).}
\label{SEDs}
\end{figure*} 
\addtocounter{figure}{-1}
\begin{figure*}
\centering 
\includegraphics[angle=270,scale=0.27]{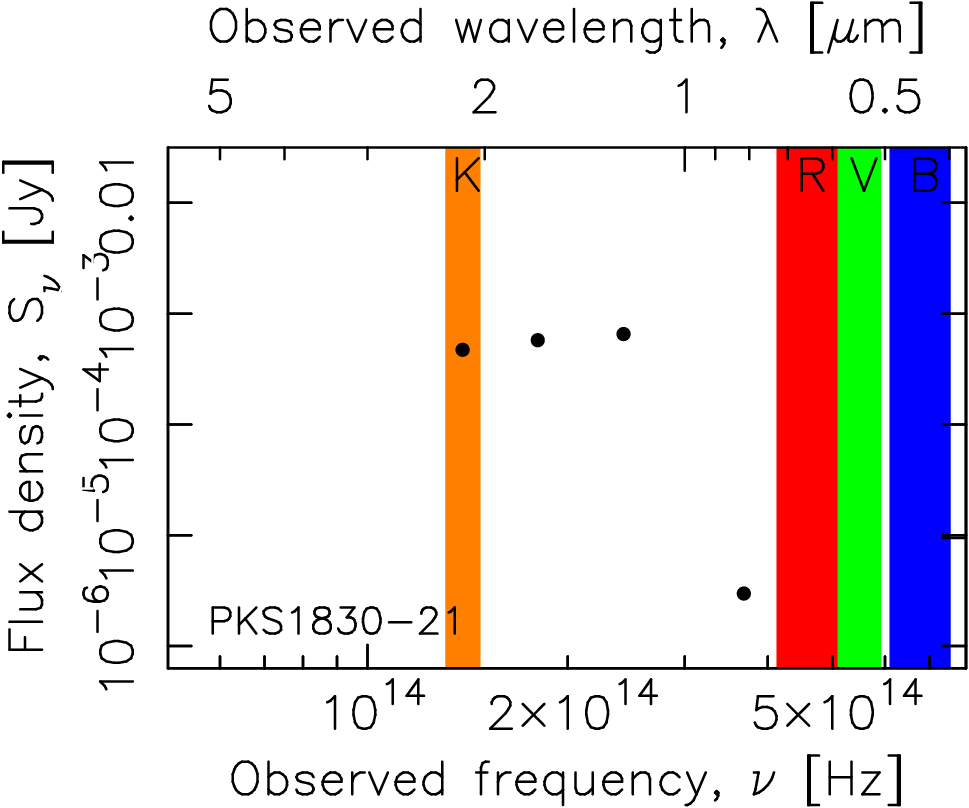}
\includegraphics[angle=270,scale=0.27]{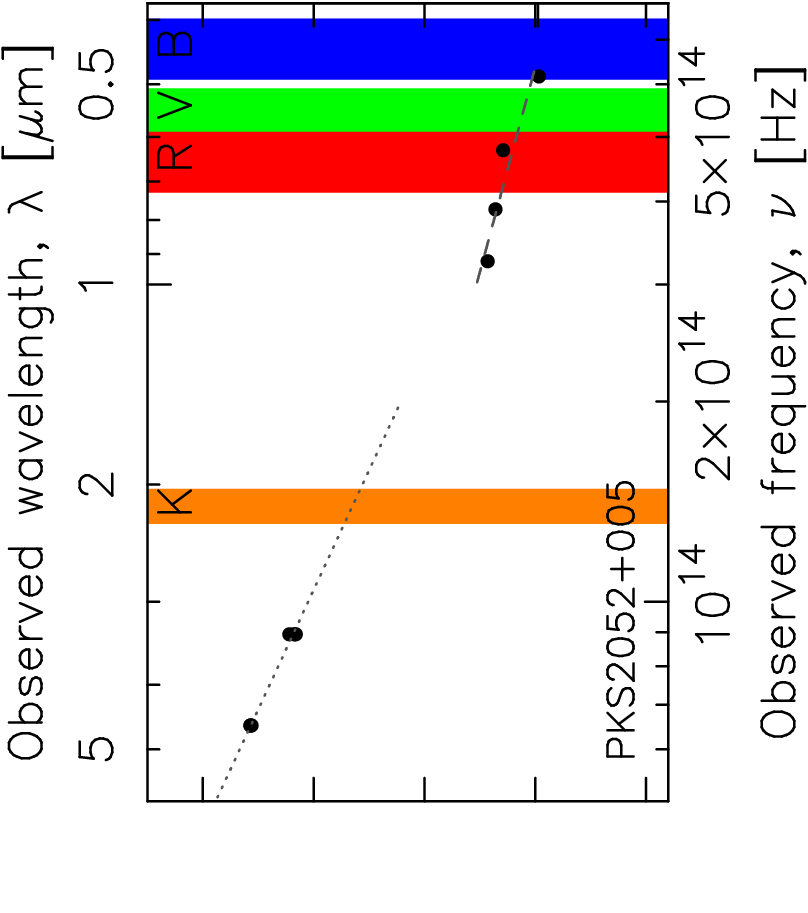}
\includegraphics[angle=270,scale=0.27]{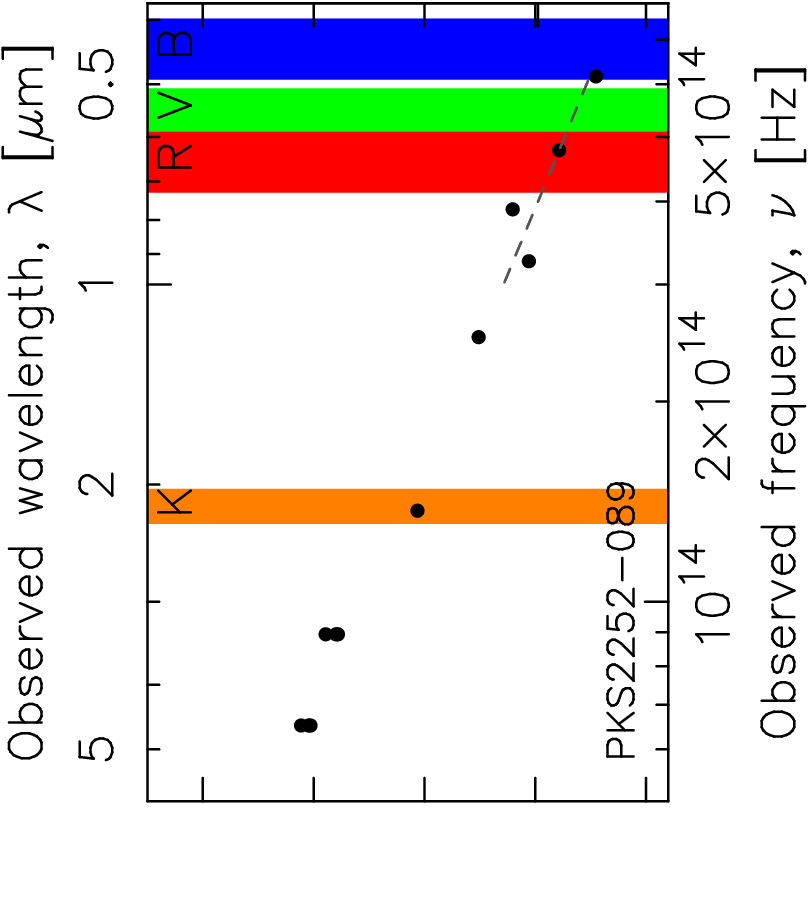} 
\includegraphics[angle=270,scale=0.27]{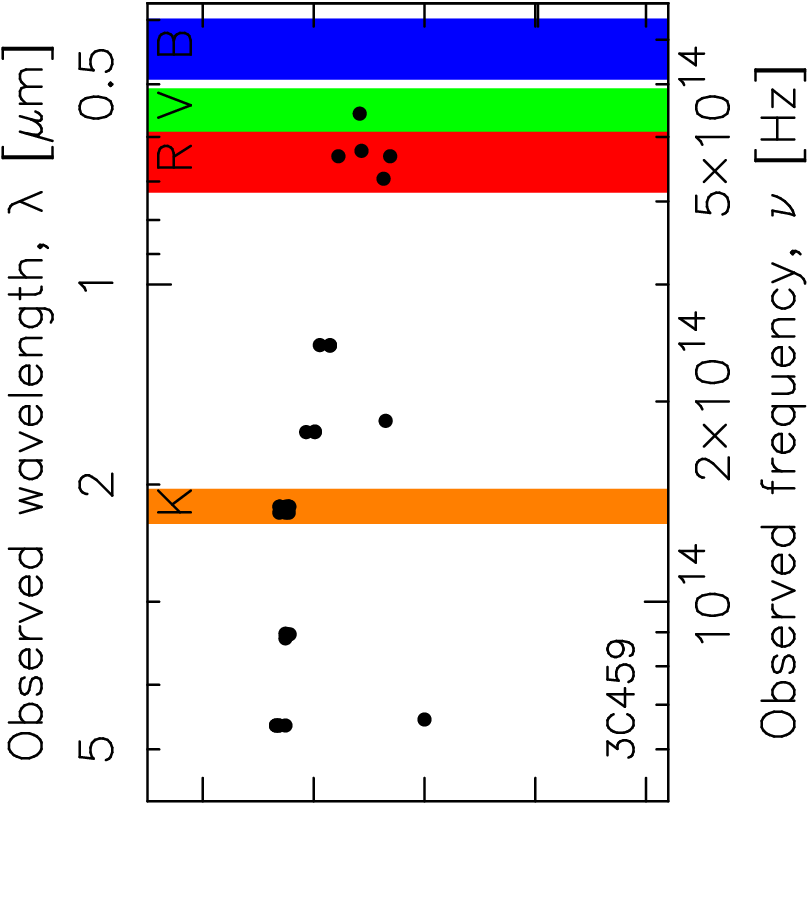}
\caption{continued.}
\end{figure*} 

Our photometry scraping does not yield a $V$ magnitude for PKS\,1830--21. Although, we had previously
(e.g. \citealt{cdbw07}) used $V-K = 6.25$, we have no record of the magnitude's origin.  Being one of the few
detections, this is important for our analysis and so on the presumption that $V>I = 22.33$, and using the 
measured $K = 15.38$, we assign $V-K
> 6.95$.  We apply the same reasoning to the other sources with a $K$ magnitude  but no $V$ measurement nor fit --- for
PKS\,0428+20  $R=20.0$, giving $V-K > 5.5$, PKS\,0500+019 $R=21.0$, giving $V-K > 5.4$ and for PKS\,1107--187 $R=19.6$, 
giving $V-K > 4.4$.  Although as yet unsearched for OH absorption, we note that PKS\,B1740--517, detected in CO $J=1\rightarrow2$ 
\citep{amm+19}, thus being the only other detected redshifted rotational-band absorber  (Sect.~\ref{sec:mm}), has $R=20.9$ giving $V-K > 5.6$.

\subsection{OH limits}
\label{sec:limits}

\begin{figure}
\centering 
\includegraphics[angle=270,scale=0.55]{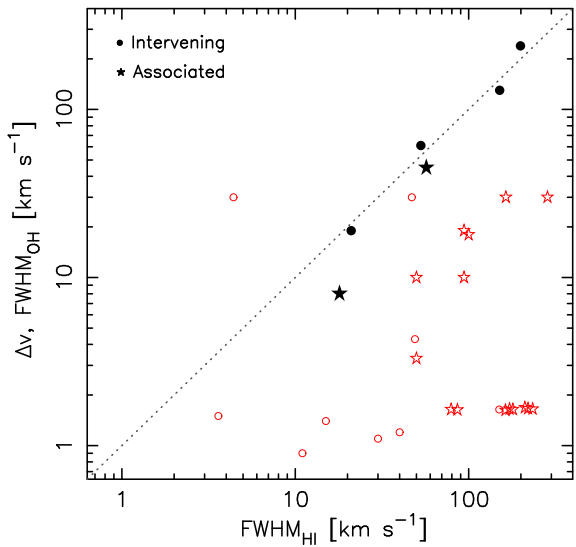}
\caption{The full width at half-maximum of the 1667-MHz OH profile versus that of the \HI\ 21-cm for the OH detected
  systems (filled symbols) and the quoted spectral resolution for the non-detections (unfilled).  The dotted
  line shows FWHM$_{\text{OH}}=\,$FWHM$_{\text{\HI}}$.}
\label{FWHM}
\end{figure}

As seen from Table~\ref{sources}, the limits to the non-detections are given at a wide variety of spectral resolutions,
therefore requiring normalisation (see \citealt{cur12}).  It was previously noted that for the OH detections the full
width at half-maxima of the profiles were very similar to those of the \HI\ \citep{cdbw07}.\footnote{As per the literature,
  when discussing OH we are referring to the 1667~MHz $^{2}\Pi_{3/2} J = 3/2$ transition.}  Showing FWHM$_{\text{OH}}$
versus FWHM$_{\text{\HI}}$ in Fig.~\ref{FWHM}, we see this is also the case for the new detection, [HB89]\,0248+430
\citep{gms+18}, with the only real outlier being the $z_{\rm abs} = 0.247$ absorber in [HB89]\,1413+135. For this,
FWHM$_{\text{\HI}} = 18$~\kms\ \citep{cps92} compared to FWHM$_{\text{OH}} \approx8$~\kms, which we obtain from our own
fit as only the total width of $\approx14$~\kms\ is quoted \citep{kc02a}.
With an rms noise level of 1.6~mJy per $\Delta v =1.75$~\kms\ and a peak of $7.8$~mJy, from our fitting, this is a weak 
detection making the line width quite uncertain. Therefore, 
we have no reason to believe that FWHM$_{\text{OH}}\not\approx\,$FWHM$_{\text{\HI}}$.  As per
\citet{cww+08,cwm+10}, we rescale the OH limits by $\sqrt{{\rm FWHM_{HI}}/\Delta v}$, in order to give the $3\sigma$
limit of a single channel ``smoothed'' to FWHM$_{\text{OH}}$.

\subsection{Results}

\begin{figure*}
\centering 
\includegraphics[angle=270,scale=0.57]{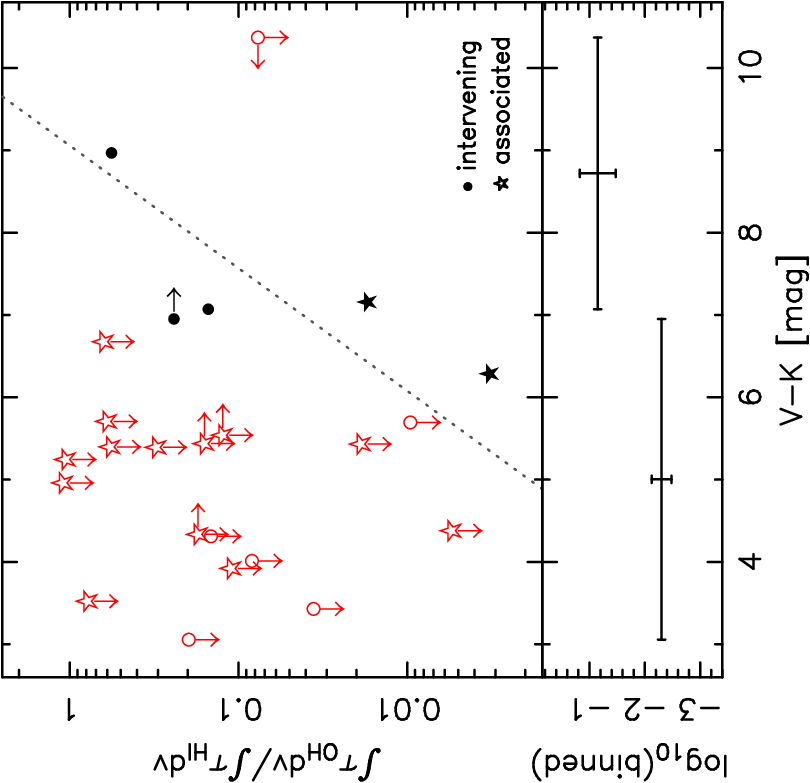}
\includegraphics[angle=270,scale=0.57]{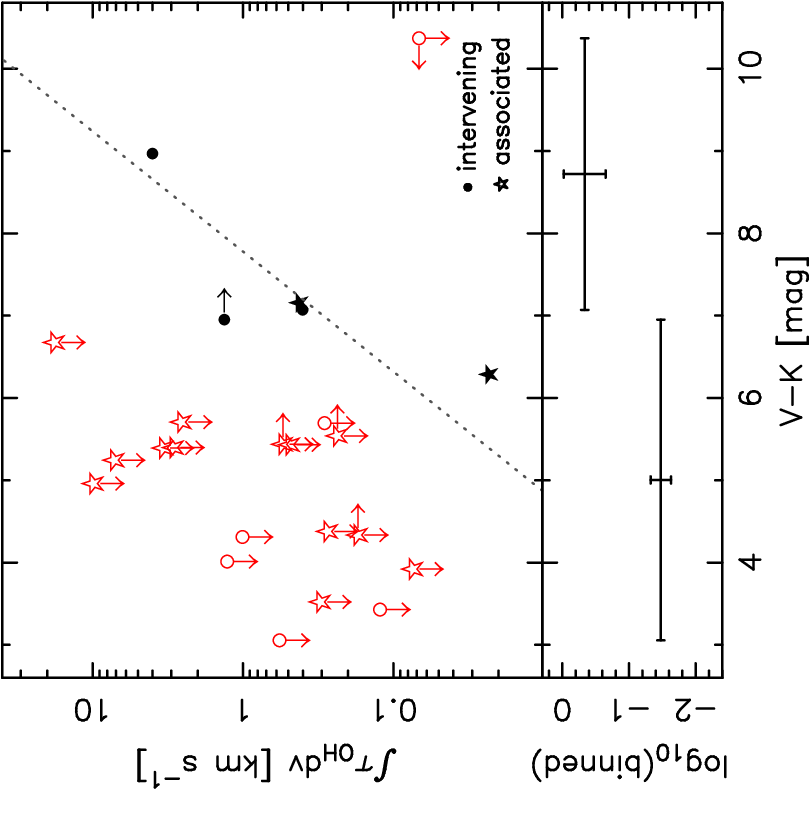}
\caption{The normalised (left) and un-normalised (right) OH line strengths for the detections (filled symbols) and the
  $3\sigma$ upper limits for the non-detections (unfilled). The dotted line shows the least-squares power-law fit to the
  five detections ([HB89]\,0248+430 is missing since it has no optical-band photometry, Fig.~\ref{SEDs}), incorporating
  the lower limit to the colour. The binned values show the upper limit to the mean ratio and the mean ratio
  $\pm1\sigma$.}
\label{VminusK}
\end{figure*}

Since our previous analysis of the relationship between the  absorption strength and $V-K$ colour \citep{cwm+10}, there 
has only been one new detection of OH \citep{gms+18}, although a plethora of non-detections
\citep{caw+16,cwa+17,gms+18,gss+21,gdb+15,gdlb20,dsgj19,zls+20}. As stated above, all OH absorbers were previously detected in
\HI\ 21-cm absorption and so we consider only these (Table~\ref{sources}).
In order to incorporate the limits (Fig.~\ref{VminusK}), we use 
the {\em Astronomy SURVival Analysis} ({\sc asurv}) package \citep{ifn86} which adds these as censored data points, allowing 
a generalised non-parametric Kendall-tau test.\footnote{Since {\sc asurv} cannot apply an upper and 
lower limit simultaneously, we do not include the three non-detections with $V-K$ lower limits (Sect.~\ref{sec:colours}).}
In Table~\ref{sig}, we summarise the two-sided probability of the observed distributions occuring by chance and the 
associated significance, assuming Gaussian statistics. Although, due to the small number of OH detections, the
correlations are not greatly significant,
\begin{table}  
\centering
 \caption{The statistical significance of the correlations, $S(\tau)$,
where $n$ is the sample size and $P(\tau)$ the probability of the distribution arising by chance.}
\begin{tabular}{@{}l  ccc  cc @{}}
\hline
\smallskip       
                       &     & \multicolumn{2}{c}{Normalised} & \multicolumn{2}{c}{Un-normalised} \\
Description & $n$ & $P(\tau)$ & $S(\tau)$ & $P(\tau)$ & $S(\tau)$ \\
\hline
With  4C\,+05.19  $V-K$  limit & 21 & 0.018 & $2.52\sigma$ &  0.0058 & $2.76\sigma$\\
No  4C\,+05.19  $V-K$  limit & 21 & 0.028 & $2.20\sigma$& 0.018 & $2.37\sigma$ \\
Detections only & 5 & 0.033&  $1.46\sigma$& 0.040&  $2.05\sigma$\\
\hline
\end{tabular}
\label{sig}  
\end{table} 
it is apparent from Fig.~\ref{VminusK} that all but one of \HI\ absorbers searched for OH absorption have
not reached the required sensitivity based upon the degree of reddening. 
For the only source which has been, 4C\,+05.19 (J0414+054), there is much debate over the origin of its extremely red
colour ($V-K =10.37$), whether it arises in the intervening gravitational lens responsible for the \HI\ absorption
\citep{cdbw07}, the $z_{\rm em} = 2.639$ background emitter \citep{fls97,tk99}, which also exhibits strong \HI\ 21-cm
absorption \citep{mcm98}, or another possible foreground object \citep{eht+06,tcw+13}. We therefore flag the
optical--near-infrared colour as $V-K < 10.37$.  Using this upper limit and the $V-K$ lower limit to PKS\,1830--21
(Sect.~\ref{sec:colours}), results in a significance of $S(\tau)=2.52\sigma$ for the normalised OH line-strength and
$2.76\sigma$ for the un-normalised line-strength.
These fall to $2.20 \text{ \& } 2.37\sigma$, respectively,  when $V-K =10.37$ is used  and $1.46\text{ \& } 2.05\sigma$, respectively,  for the five detections with $V-K$ 
measurements only.

In addition to 4C\,+05.19, for one other source the normalised OH strength appears close to detection (Fig.~\ref{VminusK}, left) -- 
PKS\,2052+005, which has  $V-K = 5.69$ and $\int\tau_{\text{OH}}dv/\int\tau_{\text{\HI}} dv < 0.01$. This limit is based upon the 
rms noise level of $\sigma_{\rm rms} = 0.6$~mJy per $\Delta v = 1.65$~\kms\ after removal of the bandpass
ripple, although RFI still affects the bandpass close to the expected absorption frequency \citep{zls+20}.

In the bottom panels of Fig.~\ref{VminusK}, we show the binned values,
where the upper limits are included via the Kaplan-Meier estimator \citep{km58}. This gives a maximum
likelihood estimate based upon the parent population \citep{fn85} and requires at least one detection in each bin,
limiting the number of bins to two.  From this, it is clear that, even without the $V-K$ limits\footnote{It is not
  possible to simultaneously include the $V-K$ limits in the binning, but we can see how the normalised OH absorption
  strength varies between the two $V-K$ bins.}, there is a positive correlation between the OH absorption 
strength and the  degree of reddening.

\section{Discussion}

\subsection{Depth of current OH searches}

By adding all of the published searched for OH absorption to date and formalising the measurement of the
optical--near-infrared colour, we confirm our previous assertion that the OH absorption strength is
correlated with the $V-K$ colour \citep{cwm+06}. This has a physical motivation, since we expect higher molecular
abundances along dustier sight-lines, which would be evident through a higher degree of reddening.  Nevertheless, this
is usually not a consideration when discussing the scarcity of redshifted OH absorption (e.g. \citealt{gms+18,zls+20}).
Here we show that, for the sensitivities reached by these surveys, the sources are simply not sufficiently red, which is
most likely due to a bias towards the most optically bright sources, in which spectroscopic redshifts are more readily
available (e.g. in DLAs). For instance, \citet{gms+18} note that the absorber with the highest \HI\ column
density\footnote{Actually integrated optical depth (see \citealt{cag+13}).} in their sample, 2MASS\,J09422198+0623362, 
exhibits no OH absorption at a $3\sigma$ limit of $\int\tau_{\rm OH} dv < 0.07$~\kms\ per 3.3~\kms\ channel. 
This has  $V-K = 4.38$ and, from the normalised fit  (Fig.~\ref{VminusK}, left),
\begin{equation}
\int\tau_{\rm OH} dv = \int\tau_{\text{\HI}}dv\times10^{0.671(V-K) - 6.078},
\label{Eq1}
\end{equation}
where $\int\tau_{\text{\HI}} dv = 49.9$~\kms, we expect $\int\tau_{\rm OH} dv=0.036$~\kms\ and,
from the un-normalised fit (Fig.~\ref{VminusK}, right),  
\begin{equation}
\int\tau_{\rm OH} dv = 10^{0.685(V-K) - 5.328},
\label{Eq2}
\end{equation}
we expect $\int\tau_{\rm OH} dv=0.005$~\kms.
Resampling the $3\sigma$ limit  to FWHM$_{\text{\HI}}$, gives $\int\tau_{\rm OH} dv < 0.3$~\kms\ per $\Delta v =50$~\kms,
which is significantly higher than either estimate thus explaining the non-detection. 

For the only \HI\ absorber which has been searched sufficiently deeply, but not detected in OH,
there has been some debate over the source of the reddening, whether in the $z_{\rm abs} = 0.96$ gravitational lens, the
$z_{\rm em} = 2.64$ host galaxy, or an another intervening galaxy (Sect.~\ref{sec:limits}). We note that the \HI\ 21-cm
absorption in the host galaxy is five times stronger than in the lens ($\int\tau_{\text{\HI}} dv = 4.2$~\kms,
\citealt{mcm98}, cf. $0.9$~\kms, \citealt{cdbw07}) and, since the 21-cm absorption strength is also correlated with
$V-K$ \citep{chj+19}, we may expect the bulk of the reddening to occur in the host galaxy. However, \citet{mcm98} did
not detect HCN $J=0\rightarrow1$ absorption to a $3\sigma$ optical depth limit of $\int\tau_{\text{HCN}} dv <
0.05$~\kms\ at $\Delta v = 19$~\kms.  It should be noted, however, that, although there is strong \HI\ and OH absorption
in the $z_{\rm abs} = 0.76$ absorber towards PKS\,0132--097, HCO$^+$ $J=1\rightarrow2$ was undetected at a $3\sigma$
limit of $\int\tau_{\text{HCO}^+}dv < 0.03$ \kms\ per 5~\kms\ channel \citep{kcl+05}. This was subsequently
detected with $\int\tau_{\text{HCO}^+}dv = 42$ \kms\ \citep{wck18} and so it is possible that  4C\,+05.19 could 
also be subject to the same apparent
variability. In any case,  the non-detection of HCN could be due to the smaller filling factor of the millimetre-band
molecular absorption \citep{zp06}, whereas the OH is more diffuse, as evident through the profile widths tracing those
of the \HI\ (Fig.~\ref{FWHM}).

\subsection{Reddening of 4C\,+05.19}

\begin{figure}
\centering 
\includegraphics[angle=270,scale=0.52]{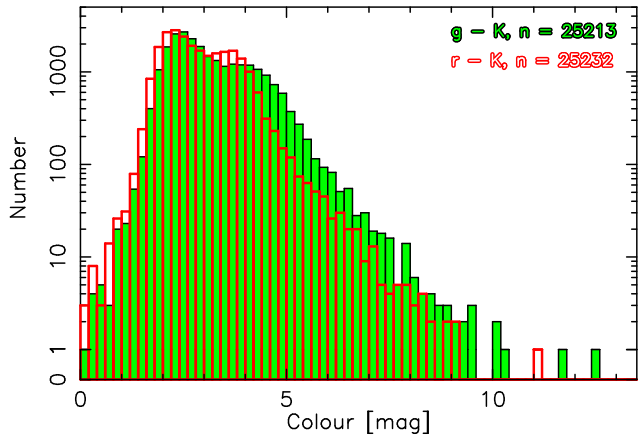}
\caption{The $g-K$ (filled histogram) and $r-K$ (unfilled histogram) colours available for the 100\,337 SDSS DR12 QSOs for which we have the photometry.}
\label{gminusK}
\end{figure}

In order to check the validity of placing an upper limit on the red colour due to the $z_{\rm abs} = 0.96$ gravitational
lens towards 4C\,+05.19, we can compare this sight-line with a large sample of quasi-stellar objects (QSOs) to gauge  just
how likely it is that the reddening is due to a single object.  For this we use the photometry compiled by \citet{cmp21}
for the first 100\,337 QSOs with accurate spectroscopic redshifts
($\delta z/z<0.01$) of the  {\em Sloan Digital Sky Survey} (SDSS) Data Release 12 (DR12, \citealt{aaa+15}). With a central 
wavelength of $\lambda = 551$~nm, the $V$-band is located between the SDSS $r$ ($\lambda = 623$~nm)
and $g$ ($\lambda = 477$~nm) bands.  We would therefore expect $r-K \lapp10.37 \lapp g-K$
for 4C\,+05.19. Only two sources satisfy this condition (Fig.~\ref{gminusK}),
SDSS\, J113415.20+392826.1 ($z_{\rm em} = 4.824$), with $g-K = 12.56$, and 
SDSS\,J150759.72+041512.1  ($z_{\rm em}  = 1.703$), with $g-K = 11.73$ (and $r-K = 11.19$). Out of
$\approx25\,000$ sources, this makes the colour of 4C\,+05.19 extremely  rare.
We note that the two SDSS sources are also at high redshift, thus being more likely to be subject to 
additional reddening due to intervening gas along the sight-line.\footnote{The obvious way to quantify  this
would be the correlation between colour and redshift, but since this is in the observed-frame it
is an inconsistent measure of the source properties over a large redshift range \citep{cm19}.} It is therefore perfectly
plausible  that all of the reddening towards 4C\,+05.19 is not solely due to the intervening lens and
that the host galaxy would be a good candidate for OH absorption.
Assuming the power-law fits to the detected OH absorbers,
we estimate the reddening due to  the $z_{\rm abs} = 0.96$ gravitational lens to be 
$V-K \leq7.4$ (Equ.~\ref{Eq1}) and  $V-K \leq6.1$ (Equ.~\ref{Eq2}), both of  which remain relatively high limits.

\subsection{Millimetre-band absorption}
\label{sec:mm}

In the millimetre (mm)-band the many rotational transitions of the large variety of molecules provide an excellent probe
of the physical and chemical conditions of the gas and its potential for star formation
(e.g. \citealt{isr88,abbj95,hjk+05}).  Sensitive temperature measurements from the relative line strengths can also
offer extremely accurate measures of the cosmic evolution of the microwave background temperature at various look-back
times \citep{wc97,mbb+12} and time delays due to molecular absorbing gravitational lenses 
can be used to measure cosmological parameters
\citep{wc01}. Furthermore, since these transitions are insensitive to the value of the fine structure constant,  but
sensitive to the electron--proton mass ratio, mm-band lines provide useful ``anchors'' to measure shifts in the line
frequencies and diagnostics of the values of the fundamental constants at large look-back times
\citep{dwbf98,cdk04}.

However, these are equally as rare, with the five OH absorbers, in conjunction with the CO $J=1\rightarrow2$ absorption
in the $z=0.44$ radio galaxy PKS\,B1740--517 \citep{amm+19}, constituting all of the redshifted mm-band
absorbers. There has also been a number of non-detections of the strongest transitions (CO, HCO$^+$ \& HCN)\footnote{See
  \citet{cw97,mbg+11,mmb+16,mmb+17,wmr+19,tcc+20} for the other detected transitions.}  published
\citep{wc95,wc96b,dcw96,mcw03,cww+08,cwc+11}\footnote{57 transitions along 43 different sight-lines.}, the majority of
which have been searched to sufficiently deep sensitivities based upon their $V-K$ colour \citep{cwc+11}. However,
applying the same prerequisite as for the OH searches, only three, [HB89]\,0438--436, PKS\,0500+019 \& PKS\,1107--187,
have been detected in \HI\ 21-cm absorption.\footnote{Although several are DLAs, i.e. strongly detected in \HI\
  Lyman-$\alpha$ absorption.} The latter two belong to the OH sample (Table~\ref{sources}) and have been searched in
HCO$^+$ $1\rightarrow2$ and CO $0\rightarrow1$, respectively \citep{cwc+11}, with [HB89]\,0438--436 having been 
searched in CO $2\rightarrow3$ \citep{dcw96}.

\begin{figure*}
\centering 
\includegraphics[angle=270,scale=0.60]{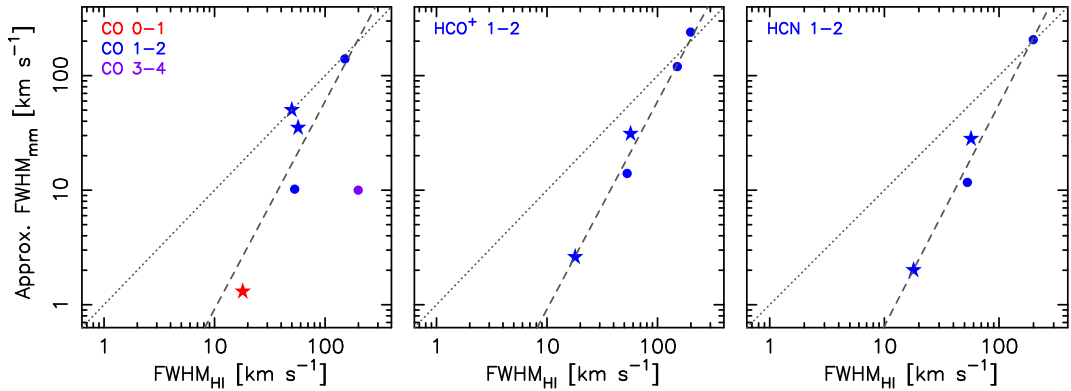}
\caption{The full width at half-maximum of the CO (left), HCO$^+$ (middle) and HCN (right) versus that of the \HI\ 
21-cm for the millimetre-band detected systems. The dotted line shows FWHM$_{\text{mm}}=\,$FWHM$_{\text{\HI}}$ and
the dashed lines the least-squares fit to the HCO$^+$ (which is also used in the left panel) and HCN  $1\rightarrow2$ transitions.} 
\label{FWHM_mm}
\end{figure*}

Again, the noise levels of the non-detections are quoted at various spectral resolutions (0.51 -- 3.4~\kms) and so we
must normalise each one  according to the expected line-width.  In Fig.~\ref{FWHM_mm}, we show the approximate millimetre
band profile widths\footnote{Often these are not given \citep{wc98,wc96,wck18,mbb+12} and have to be estimated from
  FWHM$\,\approx\int\tau dv/\tau_{\rm peak}$  (e.g. \citealt{acsr13}).} versus that of the \HI\ 21-cm.
From this we see that, in general, FWHM$_{\text{mm}}\not\approx\,$FWHM$_{\text{\HI}}$, with the divergence decreasing
with increasing line width. Although we fit the data with a power law, we suspect that the \HI\ and mm line widths
remain equal at FWHM$_{\text{\HI}}\gapp100$~\kms.  This non-equivalence could be due to the dense mm-band absorbing gas
being more localised than a more diffuse OH, although this requires the OH to be non-coincident with the other molecular
species.  Alternatively, \citet{cdbw07} suggested that the non-detection of mm-band absorption in 4C\,+05.19 (and
PKS\,0132--097) could be due to the millimetre wavelength emission, as opposed to or in addition to the absorption,
being much more localised than that at decimetre wavelengths, so that a molecular cloud has a much smaller chance of
occulting the higher frequency radiation.  However, as seen in Fig.~\ref{FWHM_mm},
FWHM$_{\text{mm}}\rightarrow\,$FWHM$_{\text{\HI}}$ with increasing line width, which would suggest that these have
correspondingly larger mm-emission regions.  While high resolution imaging in the millimetre-band is available for the
two broadest absorbers, PKS\,0132--097 and PKS\,1830--21, revealing the lensed components (\citealt{wck18,mjhm20},
respectively), imaging of the remaining sight-lines would be required to confirm this.

\begin{figure}
\centering 
\includegraphics[angle=270,scale=0.55]{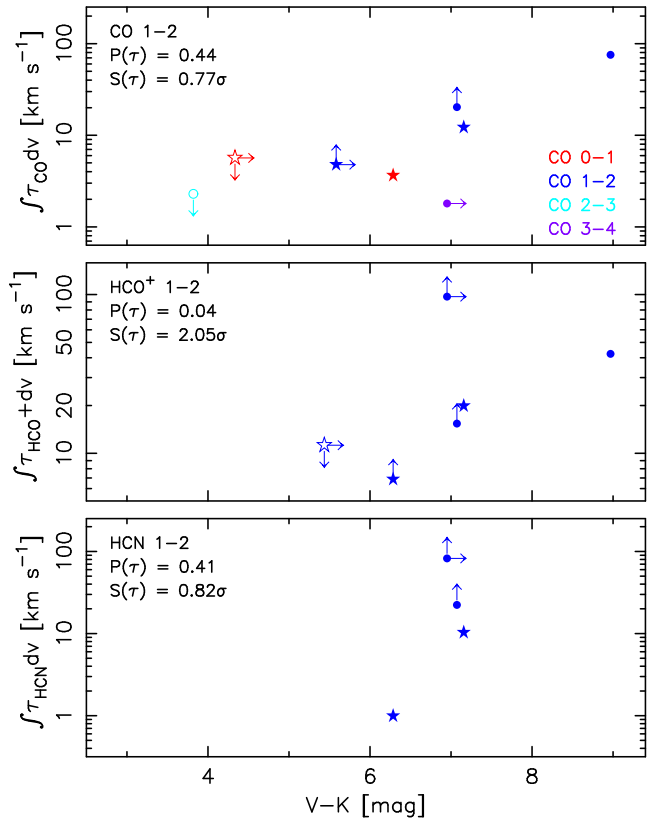}
\caption{The CO (top), HCO$^+$ (middle) and HCN (bottom) line strengths for the rotational-band absorbers. 
The statistics are shown for the $J=1\rightarrow2$ detections (filled symbols).
Note the relative weakness of the CO $3\rightarrow4$ transition, indicating the low excitation temperature ($T_{\rm ex}\sim10$~K, \citealt{wc98}).} 
 \label{mm}
\end{figure}

Resampling the limits to the FWHM expected from the HCO$^+$ $1\rightarrow2$ transition\footnote{The CO data could not be fit and the smaller sample of HCN data have a similar fit.}, in Fig.~\ref{mm}, we show the
limits together with the absorption strengths of the most commonly detected transitions versus the
optical--near-infrared colour.
Note that some of the detections are lower limits, since the millimetre-band absorption is often
optically thick, and so the assumption that the optical depth, $\tau$,  is the ratio of the observed optical 
depth $\tau_{\rm obs}$ to the
fraction of flux intercepted, $f$, cannot be made. That is,
\[ 
\tau \equiv-\ln\left(1-\frac{\tau_{\rm obs}}{f}\right) \not\approx  \frac{\tau_{\rm obs}}{f}.
\]
Due to the lower limits, we cannot obtain a least-squares fit  and, although the correlations are weak due to the very
limited numbers, an increase in $\int\tau_{\text{mm}}dv$ (and $\int\tau_{\text{mm}}dv/\int\tau_{\text{\HI}} dv$) with $V-K$ is evident. 
From this it is
clear that, although many sources have been searched to sufficiently deep limits on the basis of their colour alone
\citep{cwc+11}, those few which have also been detected in \HI\ 21-cm fall far short of the required
sensitivities. This re-emphasises the need to target extremely red sight-lines ($V-K \gapp6$), which have 
previously been detected in 21-cm absorption.

\section{Conclusions}

Since the suggestion that the strength of redshifted OH absorption is correlated with the
optical--near-infrared colour of the sight-line \citep{cwm+06}, there have been many unsuccessful searches for OH
absorption.  However, these studies do not take the red colour into account, leading to some puzzlement with regard to
the continual non-detections (e.g. \citealt{gms+18}). Here we add the new to the previous searches and, taking into
account that all of the six known OH absorbers were previously detected in \HI\ 21-cm absorption: 
\begin{enumerate}
\item In order to be applicable to future searches with the SKA and its pathfinders, rather than compiling the
  magnitudes from the variety of disparate sources in the literature (e.g. \citealt{cww+08}), we automate and formalise the 
  attainment of the $V-K$ colour by:
\begin{enumerate}
\item Scraping the photometry from NED, 2MASS and WISE, extracting the $V$ and $K$ magnitudes from the 
former two.
\item Where these are not available, extrapolating from near-by bands (WISE $W1$ \& $W2$ for the mid-infrared and $R$ \& $B$ for
the visible).
\item  And, where only a single optical-band photometry datum exists which is outside, but close to, the $V$-band, we 
use this to place a limit on the $V$ magnitude. 
\end{enumerate}
\item In order to normalise the limits,  since FWHM$_{\text{OH}}\approx\,$FWHM$_{\text{\HI}}$, we re-sample the limits to FWHM$_{\text{\HI}}$,  giving the $3\sigma$ limit of a single channel smoothed to FWHM$_{\text{OH}}$.
\end{enumerate}
From the resulting OH absorption strengths we find 
that, according to their $V-K$ colour, only one of the published limits is sufficiently
sensitive to detect OH absorption. 
This is in the $z_{\rm abs} = 0.96$ gravitational lens towards 4C\,+05.19 \citep{cdbw07}. However, this sight-line
contains at least one other \HI\ 21-cm absorber \citep{mcm98} and the source of its extremely red colour ($V-K =
10.37$) has long been a subject of debate \citep{fls97,tk99,eht+06}.  Comparing the colour with those of a large sample
of QSOs, we find this to be at the tail of the $V-K$ distribution ($\approx8\times10^{-5}$ of the sample). Using the
$V-K$ fits to the known OH absorbers, we estimate the colour due to  the lens to be 
$V-K \leq 6.1$ (based upon the un-normalised   $\int\tau_{\rm OH}dv$ limit ) and $V-K \leq 7.4$ 
($\int\tau_{\rm OH}dv/\int\tau_{\text{\HI}} dv$ limit). Given that the 21-cm
absorption in the $z_{\rm em} = 2.639$ continuum source is five times stronger than in the lens, suggests
that the host quasar is likely to contribute significantly to the degree of reddening.

Using $V-K < 10.37$ for 4C\,+05.19 and adding the limits as censored data points, we find that the 
observed $\int\tau_{\text{\HI}}dv/\int\tau_{\rm OH}dv$---$V-K$ and 
$\int\tau_{\rm OH}dv$---$V-K$  distributions to be significant at $S(\tau) = 2.52\sigma$
and $2.76\sigma$, respectively. Although not highly significant, 
due to the low number
of non-censored data points (five of the detections), binning of the data shows a clear correlation between
the normalised OH strength and the $V-K$ colour of the sight-line, thus confirming that 
the degree of reddening is a key contributor in the detection of molecular absorption.

Since five of the redshifted OH absorbers are the same sources in which the equally elusive mm-band
absorbers have been detected, we apply the same prerequisite of \HI\ 21-cm absorption. 
Although several sources have been
searched in CO and HCO$^+$ absorption to sufficiently deep limits based upon their $V-K$ colour \citep{cwc+11}, those
three which have been detected in \HI\ 21-cm have not.  Thus, we demonstrate that the degree of reddening is also
crucial in the detection of the rotational transitions. 
 This could explain the severe dearth in redshifted mm-band absorption
(also limited to six detections), with previous searches having been biased towards sources which are sufficiently bright in the
optical-bands (e.g. damped Lyman-$\alpha$ absorbers) to yield a redshift, to which to tune the narrow band millimetre receivers.
As per the OH, we therefore suggest that sight-lines with $V-K\gapp6$ detected in 21-cm absorption should be
targetted in order to increase the number of known molecular absorbers in both the millimetre and decimetre-bands.

\section*{Acknowledgements}
I wish to thank the referee for their helpful comments.
This research has made use of the NASA/IPAC Extragalactic Database (NED) which is operated by the Jet Propulsion Laboratory, California
Institute of Technology, under contract with the National Aeronautics and Space Administration and NASA's Astrophysics
Data System Bibliographic Service. This research has also made use of NASA's Astrophysics Data System Bibliographic Service
and {\sc asurv} Rev 1.2 \citep{lif92a}, which implements the methods presented in
\citet{ifn86}.

\section*{Data availability} 

Data available on request.


\label{lastpage}

\end{document}